\documentclass[8pt]{article}
\usepackage[english]{babel}
\usepackage[letterpaper,top=2cm,bottom=2cm,left=2cm,right=2cm,marginparwidth=1.25cm]{geometry}
\usepackage{amsmath,amssymb}
\usepackage{orcidlink}
\usepackage{physics}  
\usepackage{braket}  
\usepackage{graphicx}
\usepackage{tabularx}
\usepackage{float}
\usepackage{setspace}
\usepackage{xcolor}
\usepackage{comment}
\usepackage{caption}
\usepackage{subcaption}
\usepackage{soul,ulem}
\usepackage{lipsum}
\usepackage{authblk}

\title{Probing information theoretic measures of nonlinear ultracold quantum gases using phase-space distributions}
\author[1]{Mariyah~Ughradar\orcidlink{0009-0009-0554-8968}\thanks{ds22ph003@phy.svnit.ac.in
}}
\author[2]{Ramkumar~Radhakrishnan \orcidlink{0000-0001-6838-9153} \thanks{rradhak2@ncsu.edu}}
\author[3]{Siddharth~Kumar~Tiwari \orcidlink{0000-0002-3282-1028}\thanks{sidkt2015@gmail.com}}
\author[1]{Vikash~Kumar~Ojha \orcidlink{0000-0002-0641-4015} \thanks{vko@phy.svnit.ac.in}}

\affil[1]{Department of Physics, Sardar Vallabhbhai National Institute of Technology, Surat 395007, India}
\affil[2]{Department of Physics and Astronomy, North Carolina State University, Raleigh, NC 27695, USA}
\affil[3]{Shailesh J.\ Mehta School of Management, Indian Institute of Technology Bombay, Powai, Mumbai 400076, India}

\date{\today}

\begin{document}
\maketitle
\begin{abstract}
We use phase space distributions, specifically the Wigner and Husimi quasi probability distributions, to study harmonically trapped Bose--Einstein condensate described by the Gross Pitaevskii equation. From the mean field ground state wavefunction we construct both distributions and their position and
momentum space marginals and we use these to compute a comprehensive set of information theoretic measures: Shannon, Wehrl, and R\'enyi entropies; Fisher information; cumulative and cross cumulative
residual entropies; mutual information; and Kullback--Leibler, Jeffreys, Cauchy Schwarz, and R\'enyi divergences. Studying these quantities as a function of the $s$-wave scattering length for a representative Rb-85 condensate, we find that stronger repulsive interactions drive increased phase space delocalization, seen by
a monotonic growth of Shannon and Wehrl entropies, while the Fisher information shows the complementary trend --- increasing in position space and decreasing in momentum space in a manner consistent with the global Fisher uncertainty bound. R\'enyi entropies and divergence
measures further reveal a systematic suppression of non classical interference and a shift toward more classical phase space structure in moving from the Wigner to the Husimi representation, with Wigner and Husimi based mutual informations converging at larger interaction
strength. We note that, because the Gross Pitaevskii framework treats the many body state as a mean field product, the mutual information computed here quantifies statistical dependence between the conjugate phase space variables of the effective one body distribution rather than genuine particle particle entanglement.
\end{abstract}
\maketitle 
\section{Introduction}
Understanding the dynamics of strongly interacting quantum systems remains a challenge across several areas of physics. Ultracold quantum gases provide a unique platform to explore the properties of complex quantum many body systems. By using optical and magnetic trapping techniques, these gases can be confined in harmonic potentials, where the external confinement plays an important role in determining the system’s collective dynamics, equilibrium shapes, excitation spectra and coherence properties. The harmonic trap not only stabilizes the gas but also introduces characteristic length and energy scales that influence its phase space representation and dynamical evolution. As a result, trapped ultra cold gases serve as an ideal environment for exploring quantum statistical effects, hydrodynamic behavior ~\cite{jeon2015introduction} and emergent collective phases ~\cite{mendoncca2008collective} using experimentally accessible observables. We note, as broader context, that phase space distribution functions play a unifying role across vastly different energy scales in physics. In high energy nuclear physics, Wigner type correlators describe gluonic matter in the Color Glass Condensate framework and characterize mode occupation and coherence in the early time quark-gluon plasma, where observables such as anisotropic flow and hydrodynamic attractors emerge from the underlying phase space structure ~\cite{phenix2010azimuthal,martinez2012boost,voloshin2002anisotropic,mukherjee2015wigner,mukherjee2014quark,giorgini2008theory,bloch2008many,leggett2001bose,dalfovo1999theory,gross2017quantum,bloch2012quantum,kar2026thermodynamic}. While the present work operates at entirely different energy and length scales and does not develop this connection quantitatively, it is worth noting that the mathematical framework of phase space distributions and information-theoretic measures is shared across these domains. The primary motivation of the present work is more immediate: ultracold atomic gases in harmonic traps provide a clean, experimentally controllable platform for studying how interactions reshape phase space structure and quantum coherence, and the information theoretic measures we compute provide a systematic and physically transparent diagnostic of this restructuring

In parallel, the use of quantum information-theoretic measures has become increasingly valuable for characterizing correlations, coherence and complexity in quantum many-body systems ~\cite{ojha2025quantum,augusiak2012many,willby2025quantum}. Quantities such as entropies ~\cite{carter2007introduction,gray2011entropy,bromiley2004shannon}, mutual information ~\cite{veyrat2009mutual,kraskov2004estimating,behera2025mutual}, divergences ~\cite{csiszar2004information,lin2002divergence,contreras2022renyi} and Fisher information ~\cite{ojha2025quantum,fisher1,falaye2016fisher} provide an understanding beyond correlation functions, allowing us to know how information is shared between degrees of freedom, how interactions affect coherence and how far the system lies from equilibrium. When applied to phase space representations, these measures provide a comprehensive characterization of both quantum and semiclassical features of the dynamics. Some topics of nonlinear optics can be also studied using the same formalism within the regime of information theory ~\cite{kirby2015feasibility,rand2016lectures}. Nonlinear optics is now an active field of both basic research and applications and can be understood using single atoms in cavities, atomic ensembles, or interatomic interactions ~\cite{chang2014quantum,rolston2002nonlinear}. Unlike purely linear systems, nonlinear optical systems can perform nonlinear tasks such as quantum state generation and photonic logic. We note at the outset an important limitation of our framework. The Gross-Pitaevskii (GP) equation is a mean field theory in which the many body wavefunction is approximated as a pure product state. As a consequence, all information theoretic measures computed in this work are derived from an effective single-particle distribution and reflect interaction-induced restructuring of the one-body phase space rather than genuine many-body entanglement or beyond mean field correlations. When we refer to 'correlations' or 'collective behavior' in what follows, we mean statistical dependences encoded in the phase space distribution of the mean field wavefunction, not inter-particle entanglement. True many-body correlations, such as those captured by beyond mean field methods including Bogoliubov theory or exact diagonalization, are outside the scope of the present work. The GP framework is nevertheless an appropriate and widely used starting point for studying phase space structure in weakly interacting Bose Einstein condensates and the information theoretic measures we compute provide physically meaningful diagnostics of interaction induced delocalization and coherence loss within this approximation.

The phase space representation is formalized through the Wigner distribution, which provides a semiclassical description of quantum states while retaining full quantum coherence ~\cite{wigner1932quantum,radhakrishnan2022wigner}. However, the Wigner function is not positive definite and can exhibit negative regions that show quantum interference and nonclassical correlations ~\cite{hudson1974wigner,kenfack2004negativity}. To obtain a strictly positive phase space representation, one can employ the Husimi distribution, a Gaussian smoothed version of the Wigner function that retains key dynamical information while eliminating negativity ~\cite{takahashi1986wigner,takahashi1985chaos,radhakrishnan2022wigner}. The Wigner distribution can show quantum features and interference, while the Husimi distribution smooths these out, giving a more classical looking picture to analyze.

 We use both Wigner and Husimi phase space distributions to study correlation structure and information flow in interacting ultracold atomic gases. To proceed, we carry out a numerical study of a bosonic quantum gas confined in a harmonic trap, described using the Gross–Pitaevskii equation ~\cite{gross1961structure,pitaevskii1961vortex} and some of the works along this line is carried out in the past (see refs. ~\cite{zhao2019optical,chakrabarti2024quantum}). From each distribution, we compute marginal densities and evaluate information theoretic measures, including Shannon and Rényi entropies, survival functions, mutual information, divergence measures (Kullback–Leibler, Jeffreys and Rényi) and the Fisher information. These measures describes phase space localization, redistribution of occupation among momentum modes, non thermal correlations and the development of coherence or mixing. We then perform a detailed comparative study of the Wigner and Husimi based measures, identifying which features are sensitive to purely quantum coherence.

The article follows this structure. In Section \ref{sec:phasespace}, we introduce phase space distributions and their associated entropies, specifically discussing the Wigner and Husimi distributions. In Section \ref{sec:infotheory}, we cover various aspects of information theory, including divergences, mutual information, and relative entropy. In Section \ref{sec:becgases}, we apply these concepts to nonlinear ultracold atomic gases and interpret the results. Definitions of all key concepts are provided in the following section prior to the discussion of results. We set $\hbar = c = 1$ throughout the article. Unless otherwise stated, all integrals are taken over the limits $[-\infty, \infty]$.
\section{Phase space distributions and Entropies}\label{sec:phasespace}
The Wigner distribution $(\mathcal{W}(x,k))$ can be defined in terms of  Wigner–Weyl transform ~\cite{case2008wigner} as
\begin{equation}\label{eq:WD}
    \mathcal{W}(x,k) = \frac{1}{\pi}\int dy\, e^{-2iky}\, \rho(x-y,x+y),
\end{equation}
where $\rho(x-y,x+y)$ is the density matrix. For a pure state, the density matrices can be expressed in terms of its wave function as
\begin{equation}
    \mathcal{W}(x,k) = \frac{1}{\pi} \int dy\,e^{-2iky}\, \psi^{*}(x-y)\psi(x+y).
\end{equation}
which satisfies the normalization property,
\begin{equation}
    \int\,\int\,dx\,dk\,\mathcal{W}(x,k) = 1.
\end{equation}
Wigner distribution is also referred to as a quasi-probability distribution because it retains many of the features of a true probability distribution . The marginals of the Wigner distribution play a central role in establishing its connection to measurable quantities~\cite{radhakrishnan2022wigner,ojha2025phase}. Despite the fact that $\mathcal{W}(x,k)$ is not a true probability distribution, since it can take negative values and it reproduces the correct quantum probability densities when integrated over either position or momentum. First, integrating over momentum yields the position space probability density:
\begin{align}
    \rho(x) & = \int\, dk\, \mathcal{W}(x,k) = |\Psi(x)|^{2}.
\end{align}
This shows that the Wigner distribution correctly describes the spatial probability density associated with the quantum state $\Psi(x)$. Similarly, integrating over position gives the momentum space probability density:
\begin{align}
    \rho(k) & = \int\, dx\, \mathcal{W}(x,k) = |\Phi(k)|^{2}.
\end{align}
where $\Phi(k)$ is the Fourier transform of the wavefunction. Thus, the Wigner distribution also reproduces the correct momentum distribution. What is nontrivial here is that both marginals are strictly non negative and properly normalized, even though the full phase space distribution $\mathcal{W}(x,k)$ can become negative in certain regions. These negative regions are not artifacts and they describes genuinely quantum features such as interference and non classical correlations.

The expectation value of any observable $(\hat{O})$  can be obtained by treating the Wigner distribution as a probability distribution, as discussed in~\cite{leonhardt2010essential}
\begin{equation}
    <\hat{\mathcal{O}}> = 2\pi \int\,\int dx\, dk\,\hat{\mathcal{O}}(x,k) \mathcal{W}(x,k).
\end{equation} 
The marginals of the Wigner distribution are positive definite. However, the Wigner distribution itself is not positive definite ~\cite{kenfack2004negativity} and can exhibit negative regions. The negative regions of the distribution reflect genuine quantum effects. Even though the distribution can take negative values, it is still properly normalizable, which is necessary if one wants to treat it like a probability distribution. However, these negative regions create difficulties when defining information theoretic quantities ~\cite{kenfack2004negativity}. For instance, the Shannon entropy of the Wigner distribution is defined as
\begin{equation}
    S_{W} =  -\int\,\int dx\,dk\,\mathcal{W}(x,k)\,\ln[\mathcal{W}(x,k)],
\end{equation}
We propose that Wigner entropy, although defined only for Wigner positive states, is a suitable measure for quantifying uncertainty in configuration space. It provides information about the uncertainties of the marginal distributions of $x$ and $k$ as well as their correlations in phase space. Unlike Wehrl entropy ~\cite{lieb1978proof}, Wigner entropy does not correspond to the classical entropy associated with a particular measurement outcome. It exhibits notable properties, including invariance under symplectic transformations in configuration space ~\cite{wehrl1979relation}. This invariance is essential for any effective measure of phase space uncertainty, as symplectic transformations preserve phase space area ~\cite{cover1991network}. In contrast, Wehrl entropy tends to be higher for squeezed states than for coherent states ~\cite{lieb1978proof}.
In the present work, the Wigner function obtained from the Gross–Pitaevskii ground state remains positive over the considered parameter range and therefore $S_{W}$ is real valued. Since the Wigner distribution contains negative regions, its corresponding entropies will be ill-defined at those negative regions. However, the Shannon entropies of the marginals of the Wigner distribution are real valued due to the positivity of the position and
momentum space densities. The entropies of the marginals quantify the entropic uncertainty relation and are given as follows:
\begin{equation}
     S_{x}^{W} = S_{x} = -\int dx\, \rho(x)\,\ln[\rho(x)],
\end{equation}
\begin{equation}
     S_{k}^{W} = S_{k} = -\int dk\, \rho(k)\,\ln[\rho(k)],
\end{equation}
where the uncertainty bound for the Shannon entropy is given by
\begin{align}
     S^{W}_{x}+S^{W}_{k} \geq 1+\ln\pi.
\end{align}
The Husimi distribution $(\mathcal{H}(x,k))$ is obtained by Gaussian smoothing of the Wigner distribution — equivalently, by a Weierstrass transform ~\cite{leonhardt2010essential,husimi1940some,harriman1988some} — and can be written as:
\begin{equation}
   \mathcal{H}(x,k) = \frac{1}{\pi}\int\,\int\,dx'\,dk'\,e^{-\frac{(x-x')^{2}}{2s^{2}}}\,e^{-(p-p')^{2}\,2s^{2}}\,\mathcal{W}(x,k), 
\end{equation}
where $s$ is an arbitrary parameter that we will set to unity in our context. The Husimi distribution is positive definite, in contrast to the Wigner distribution, which contains negative regions. Consequently, quantities derived from the Husimi distribution cannot fully reflect the system’s true characteristics ~\cite{wehrl1979relation}, as the distribution is obtained by Gaussian filtering. It also lacks certain features of the Wigner distribution, such as the overlap formula ~\cite{leonhardt2010essential} and its local structure ~\cite{appleby1999generalized}. Its Shannon entropy, known as Wehrl entropy ~\cite{wehrl1979relation}, is well defined and real valued due to the distribution’s positive nature:
\begin{equation}
    S_{H} = -\int\,\int dx\,dk\, \mathcal{H}(x,k)\ln \mathcal{H}(x,k),
\end{equation}
where $H(x,k)$ is the Husimi distribution which satisfies the normalization condition
\begin{equation}
    \int\,\int dx\,dk\, \mathcal{H}(x,k) = 1.
\end{equation}
Its marginals are 
\begin{equation}
    \rho_{H}(x) = \int dk\, \mathcal{H}(x,k) \hspace{2em},\hspace{2em}  \rho_{H}(k) =  \int dx\, \mathcal{H}(x,k).
\end{equation}
The entropies of the marginals of the Husimi distribution are
\begin{equation}
     S_{x}^{H} = -\int dx\, \rho_{H}(x)\,\ln[\rho_{H}(x)],
\end{equation}
\begin{equation}
     S_{k}^{H} = -\int dk\, \rho_{H}(k)\,\ln[\rho_{H}(k)].
\end{equation}
Another way to address the negativity in the Wigner distribution is by using cumulative or survival functions. 
\begin{equation} \label{eq:survival}
    s_{W}(c,b) = \int_{c}^{\infty} \int_{b}^{\infty}dx\,dk\,\mathcal{W}(x,k),
\end{equation}
whose survivals of the marginals are
\begin{equation}
    s_{x}^{W}(c) = \int_{c}^{\infty}dx\,\rho(x),
\end{equation} 
\begin{equation}
    s_{k}^{W}(b) = \int_{b}^{\infty}dk\,\rho(k).
\end{equation}
The areas of the distribution that are integrated correspond to regions associated with negative entropy. By applying this approach, we can define the cumulative residual entropy ($\mathcal{C}$), which is a measure used to quantify the uncertainty or randomness in a sequence of events over time ~\cite{rao2004cumulative}. It is expressed as
\begin{equation}
   \mathcal{C}_{x}^{W} = -\int\,dc\, s_{x}^{W}(c)\,\ln s_{x}^{W}(c),
\end{equation}
\begin{equation}
    \mathcal{C}_{k}^{W} = -\int db\, s_{k}^{W}(b)\,\ln s_{k}^{W}(b).
\end{equation}
An entropic uncertainty relation can be established between them. In addition, information theory provides the concept of Fisher information $(\mathcal{F})$ ~\cite{fisher1}, which quantifies the sensitivity of a probability distribution to an unknown parameter and is defined for a continuous probability density as given in Eq.~\eqref{eq:fisher33} below. In position space, for a quantum state with wavefunction $\Psi(x)$, the Fisher information takes the form$(\mathcal{F}_{x})$ is defined as (~\cite{falaye2016fisher,kharazmi2023fisher}):
\begin{equation}\label{eq:fisher33}
    \mathcal{F}_{x} = \int dx\, |\Psi(x)|^{2}\bigg[\frac{d}{dx}\ln|\Psi(x)|^{2}\bigg]^{2}   > 0, 
\end{equation}
and in momentum space
\begin{equation}\label{eq:fisher34}
    \mathcal{F}_{k} = \int dk\,|\Phi(k)|^{2}\bigg[\frac{d}{dk}\ln|\Phi(k)|^{2}\bigg]^{2}  > 0, 
\end{equation}
where Heisenberg's indeterminacy principle in terms of Fisher's information measure is formulated as
\begin{equation}
    \mathcal{F}_{x}\cdot \mathcal{F}_{k} \geq 4. 
\end{equation}
\section{Various Information theoretic measures}\label{sec:infotheory}
After obtaining the phase space distributions, we move on to calculating entropies from the perspective of information theory. As a first step, we compute the Kullback–Leibler $(S_{\text{KL}})$ divergence (relative entropy) ~\cite{van2014renyi}, which is a measure of the distance between two distributions. This can be defined in terms of the position space of Wigner and Husimi distribution marginals as
\begin{equation} \label{eq:KL}
        S_{\text{KL}} = \int dx\,\rho(x)\, \ln\left[\frac{\rho(x)}{\rho_{H}(x)}\right]. 
    \end{equation}

A complementary measure can be constructed at the level of survival
functions. We define a survival Jeffreys-type measure $\mathcal{R}$,
which acts as the survival-function analogue of the relative entropy in
Eq.~(\ref{eq:KL}), and is used to compare the cumulative tails of the
Wigner and Husimi position-space marginals:
\begin{equation}
\mathcal{R} = \int dc\, \left[
s^{W}_{r}(c)\,\ln\!\left(\frac{s^{W}_{r}(c)}{s^{H}_{r}(c)}\right)
+ s^{H}_{r}(c)\,\ln\!\left(\frac{s^{H}_{r}(c)}{s^{W}_{r}(c)}\right)
\right].
\label{eq:Rsurvival}
\end{equation}
Equivalently, $\mathcal{R}$ can be written as the symmetrized sum of
forward and reverse KL-type contributions evaluated on the survival
functions,
\begin{equation}
\mathcal{R} \equiv J_{s}(s^{W},s^{H})
= D^{s}_{\mathrm{KL}}(s^{W}\|s^{H})
+ D^{s}_{\mathrm{KL}}(s^{H}\|s^{W}),
\label{eq:Rsurvival_KL}
\end{equation}
where the superscript $s$ emphasizes that the functional is applied to
survival functions rather than to normalized probability densities.

We stress that $\mathcal{R}$ as defined in Eq.~(\ref{eq:Rsurvival}) is not the standard Jeffreys divergence between probability densities. The standard Jeffreys divergence inherits non negativity from Gibbs' inequality, which relies on the integrand being a normalized probability density. The survival functions $s^{W}_{r}$ and $s^{H}_{r}$ are non-negative and monotonically decreasing, but they are not normalized to unity over the integration domain — they take the value
of the cumulative tail probability and integrate to the mean of the underlying distribution. The Jeffreys-type symmetrization applied to non-normalized non-negative functions therefore does not, in general, inherit non-negativity, and $\mathcal{R}$ can take either sign depending on the local relation between the two survival curves. The properties that do carry over from the standard construction are symmetry under
exchange of the two survival functions and vanishing if and only if $s^{W}_{r} \equiv s^{H}_{r}$. Despite the loss of a positivity guarantee, $\mathcal{R}$ remains a useful diagnostic because it is preferentially sensitive to differences in the tails of the two distributions, where the smoothing effect of the Husimi representation
is most visible, and it therefore provides information complementary to the standard KL divergence between densities.

Having introduced relative entropy and its role in quantifying differences between probability distributions, we now turn to mutual information. Mutual information ~\cite{cover1999elements} measures how much information is shared between two random variables or sets of variables ~\cite{kraskov2004estimating}. By capturing statistical dependence or correlation, mutual information reveals key relationships within a system. Importantly, mutual information is closely related to relative entropy ~\cite{mackay2003information}, as it can be expressed in terms of it—highlighting their complementary roles in information theory. While relative entropy measures the difference between distributions, mutual information focuses on the shared information between variables. The mutual information $(\mathcal{I})$ in terms of the Wigner distribution and its marginals is given by
\begin{equation} \label{eq:mutualwigner}
    \mathcal{I}_{W} = \int\,\int dx\, dk\, \mathcal{W}(x,k)\, \ln\left[\frac{\mathcal{W}(x,k)}{
    \rho(x)\rho(k)}\right].
\end{equation}
It can also be expressed in terms of entropies from the Wigner distribution as ~\cite{salazar2023phase}
\begin{equation}
   \mathcal{I}_{W} = S^{W}_{x}+S^{W}_{k}-S_{W},
\end{equation}
where $ S^{W}_{x}+S^{W}_{k} \geq 1+\ln\pi$. We can also find mutual information using Husimi distribution (see refs. ~\cite{grabowski1984wehrl,floerchinger2021wehrl}). The analogue of mutual information in this setting of survival function is the cross cumulative residual entropy ($\mathbf{C}$) ~\cite{rao2004cumulative}. For the Wigner distribution, it is defined as
\begin{equation} \label{eq:crosscumulative}
    \mathbf{C}^{W} = \mathcal{C}^{W}_{x} -\epsilon,
\end{equation}
where $$\epsilon = -\int\,\int dk\, dx\, \int_{b}^{\infty} db\, \mathcal{W}(x,k)\,\ln\left[\frac{\int_{b}^{\infty} dk\, \mathcal{W}(x,k)}{\rho(x)}\right].$$
This phase space residual correlation measure removes purely positional survival uncertainty and isolates the contribution arising from joint position momentum structure. In the present work, mutual information is defined between the conjugate phase space variables $x$ and $k$ rather than between spatially separated particles. It is computed from the joint phase space distribution $P(x,k)$, given by either the Wigner or Husimi function, and its corresponding marginals. Here, the “subsystems” correspond to the position and momentum degrees of freedom within the reduced one-body description. Thus, the mutual information quantifies statistical dependence between these conjugate variables. Because the analysis is based on the Gross–Pitaevskii mean-field framework, this quantity does not represent genuine particle–particle entanglement. Rather, it captures interaction-induced phase space correlations encoded in the effective single-particle distribution. Rényi entropy provides a parametrized family of measures that capture the uncertainty or randomness of a probability distribution ~\cite{van2014renyi}, generalizing the Shannon entropy through the parameters $\alpha$ and $\beta$. It is given by
\begin{equation} \label{eq:RENyipos}
    \mathcal{R}_{x}^{\alpha} = \frac{1}{1-\alpha}\, \ln \left[\int dx\, \left\{[\rho(x)]^{2}\right\}^{\alpha} \right], 
\end{equation}
\begin{equation}
     \mathcal{R}_{k}^{\beta} = \frac{1}{1-\beta}\, \ln \left[\int dk\, \left\{[\rho(k)]^{2}\right\}^{\beta} \right].
\end{equation}
For the Wigner distribution, the choice  $\alpha = \beta = 2$ yields the collision entropy ~\cite{muller2013quantum}, which is given by:
\begin{equation}
    \mathcal{R}_{W} = -\ln\bigg[\int\, \int dx\, dk\, \mathcal{W}(x,k)^{2}\bigg].
\end{equation}
Even though the Wigner distribution can take negative values, the entropy remains well defined for the specific case $\alpha = 2$. The corresponding Rényi uncertainty bound ~\cite{zozor2007classes} is 
\begin{equation}
     R_{k}^{\beta} +  R_{x}^{\alpha} \geq -\frac{1}{2(1-\beta)}\ln\bigg(\frac{\beta}{\pi}\bigg)-\frac{1}{2(1-\alpha)}\ln\bigg(\frac{\alpha}{\pi}\bigg),
\end{equation}
with the restriction
\begin{equation}
    \frac{1}{\alpha} + \frac{1}{\beta} = 2.
\end{equation}
The R\'enyi divergence, also known as the R\'enyi relative entropy $S_{RD}$, quantifies the difference between two probability distributions and generalizes the Kullback--Leibler divergence through
a parameter $\alpha$ in the same way that the R\'enyi entropy generalizes the Shannon entropy. For two probability densities $\rho(x)$ and $\rho_{H}(x)$ on the same support, the R\'enyi divergence of order $\alpha$ is defined as~\cite{contreras2022renyi,van2014renyi}
\begin{equation}
S_{RD}^{(\alpha)}
= \frac{1}{\alpha - 1}\,
\ln \left[\int dx\; \rho(x)^{\alpha}\, \rho_{H}(x)^{1-\alpha}\right],
\qquad \alpha > 0,\ \alpha \neq 1.
\label{eq:Renyi_div}
\end{equation}
This expression reduces to the Kullback--Leibler divergence in the limit $\alpha \to 1$. The R\'enyi divergence is non-negative for all admissible values of $\alpha$, with $S_{RD}^{(\alpha)} \geq 0$ and equality if and only if $\rho \equiv \rho_{H}$ almost everywhere; this
follows from Jensen's inequality applied to the convex (for$\alpha > 1$) or concave (for $0 < \alpha < 1$) function $t \mapsto t^{\alpha}$, with the sign of the prefactor $1/(\alpha-1)$ compensating the sign of the inequality so that non-negativity holds in both regimes. In particular, $S_{RD}^{(\alpha)}$ cannot be negative for $\alpha < 1$.

For this particular choice of $\alpha = 2$, the quadratic Rényi mutual information ($\mathcal{I}^{2}_{W}$) is defined for the Wigner distribution as
\begin{equation}
    \mathcal{I}^{2}_{W} = \ln \bigg[\int\,\int dx\, dk\, \frac{\mathcal{W}[x,k]^{2}}{\rho(x)\,\rho(k)}\bigg],
\end{equation}
For the Husimi distribution, the corresponding quantities (Rényi entropy and mutual information) are obtained by substituting the Wigner distribution with the Husimi distribution and adjusting the probability densities in the previously introduced formulas. Another divergence, known as the Cauchy–Schwarz divergence ($\mathcal{D}_{CS}$) ~\cite{hoang2015cauchy}, has a form resembling mutual information. It quantifies the degree to which two variables or vectors deviate from perfect alignment and is specifically defined for a single variable. For given distributions $f_{1}(x)$ and $f_{2}(x)$, Cauchy Schwarz divergence is defined as 
\begin{equation}
    \mathcal{D}_{CS} = -\ln\,\Bigg[\frac{\int dx\, [f_{1}(x)f_{2}(x)]^{2}}{\sqrt{\left[\int dx\, f_{1}(x)^{2}\right]\left[\int dx\, f_{2}(x)^{2}\right]}}\Bigg].
\end{equation}
and the mutual information associated with the CS divergence is referred to as Cauchy–Schwarz mutual information.
\section{Ultracold Quantum Gases in a Harmonic Trap}\label{sec:becgases}
\subsection{Description of the system}\label{subsec:descrip}
We consider a harmonically trapped quantum gas as a representative system. One–dimensional Bose–Einstein condensates confined in harmonic traps provide a useful framework to study the interplay between quantum correlations, confinement ~\cite{williams1991organic,jerome1991physics} and interaction strength. We begin with a three-dimensional harmonic trap. By fixing the axial coordinate $z$, the problem reduces to an effective one-dimensional description, and the corresponding wave function is obtained by solving the time-dependent nonlinear Schrödinger equation (NLSE). Specifically, we employ the Gross–Pitaevskii (GP) equation ~\cite{gross1961structure,pitaevskii1961vortex}, which has the form of the Schrödinger equation with an additional interaction term. The interaction coupling is proportional to the $s$-wave scattering length of two interacting bosons. The Gross–Pitaevskii equation is given by
\begin{align}
    i\frac{\partial \Psi}{\partial t} & = \Bigg[-\frac{1}{2m}\nabla^{2}+g|\Psi(x,y,z,t)|^2+V(x,y,z)\Bigg]\Psi(x,y,z,t),
\end{align}
where $g$ is the coupling constant. We introduce the following dimensionless variables:  $$\tau = \omega_{r}t, \hspace{2em}\,a_{0}\rho = r, \hspace{2em}\,a_{0}s = z, \hspace{2em}\,Q = -\frac{8\pi\,a\, N}{a_{0}}, \hspace{2em}\,a_{0} = \frac{1}{r}\,\sqrt{\frac{1}{m\omega_{r}}}, \hspace{2em}\,\lambda_{z} = \frac{\omega_{z}}{\omega_{r}}$$
where $\omega_{r}$ is the radial trap frequency, $a$ is the $s$-wave scattering length, $\omega_{z}$ is the axial frequency and $N$ is the mean number of weakly interacting bosons. We note a brief change of notation. The general definitions in Sections~\ref{sec:phasespace} and \ref{sec:infotheory} are written in terms of generic conjugate variables
$x$ and $k$. In Section~\ref{sec:becgases} onwards, the explicit one-dimensional condensate is described in cylindrical coordinates with radial
coordinate $r$ and conjugate momentum $k_{r}$, after the axial direction has been integrated out (see Appendix~A). Throughout Section~\ref{sec:becgases}, the symbols $r$ and $k_{r}$ replace $x$ and $k$ in the
definitions of all marginal densities, survival functions, entropies and divergence measures introduced earlier. With these definitions, the condensate wave function in position space takes the form
 \begin{align}
     \Psi(r,z,t) & = \frac{i}{4\pi\, a_{0}^{2}}\sqrt{8\pi\,a\,N}\,sech\bigg(\frac{aN}{a_{0}^{2}}z\bigg)\exp{\bigg[-\frac{r^{2}}{2a_{0}^{2}}\bigg]}\,e^{i\omega_{r}t}
 \end{align}
The corresponding momentum space wave function is
 \begin{align}
     \Phi(k_{r},k_{z},t) & = \frac{i\,\pi\, a_{0}^{2}}{\sqrt{2\pi\,a N}}sech\bigg(\frac{\pi\,a_{0}^{2}}{2aN}k_{z}\bigg)\exp{\bigg[-\frac{1}{2}k^{2}_{r}a^{2}_{0}\bigg]}\,e^{i\omega_{r}t}
 \end{align}
where $a_{0}$ is the oscillator length, $a$ is the scattering length, $N$ is the mean number of trapped bosons and $\kappa = \tfrac{4\pi a N}{a_{0}}$. In this treatment, the axial direction $z$ is assumed to be fixed. The detailed derivation is presented in the Appendix~\ref{sec:appendix}.Using these wave functions, we compute the Wigner and Husimi distributions. The distributions are derived analytically, and numerical calculations are performed using the following parameter values of Rb-85 atom: $N = 10, a_{0} = 3\times 10^{-6}\, m, z = 2\times 10^{-8}\,m, a = 5 \times 10^{-10}\, m $. The parameters chosen here correspond to a dilute Rb-85 condensate in a regime that is experimentally accessible and theoretically well controlled. Rb-85 is a particularly relevant species because its s-wave scattering length is continuously tunable via magnetic Feshbach resonance across a wide range, including both positive and negative values, making it an ideal platform for studying interaction-dependent phase space structure. In the present work, we restrict ourselves to positive scattering lengths in order to ensure stability of the GP ground state: for negative scattering lengths, the mean-field interaction is attractive, and the condensate becomes unstable above a critical particle number, leading to collapse. A systematic study of the attractive regime, which would require either a stability analysis or the inclusion of beyond-mean-field corrections, is left for future work. Within the positive scattering length regime, the chosen range $a = 5\times 10^{-10}\,m$ corresponds to moderately repulsive interactions where the GP equation is well-justified and the ground state remains positive definite throughout phase space, ensuring that all entropy measures are real valued and well defined. The oscillator length $a_{0} = 3\times 10^{-6}\, m$ and the particle number $N = 10$ place the system firmly in the weakly interacting, harmonically trapped regime where the GP approximation is quantitatively reliable. We note that larger particle numbers would generically increase configurational complexity and entropic measures, as discussed in Section~\ref{sec:wigentro}, and a systematic study of the $N$-dependence is a natural extension of the present work.

\begin{figure}[t]
\centering
\begin{subfigure}{0.45\textwidth}
  \centering
  \includegraphics[width=\linewidth]{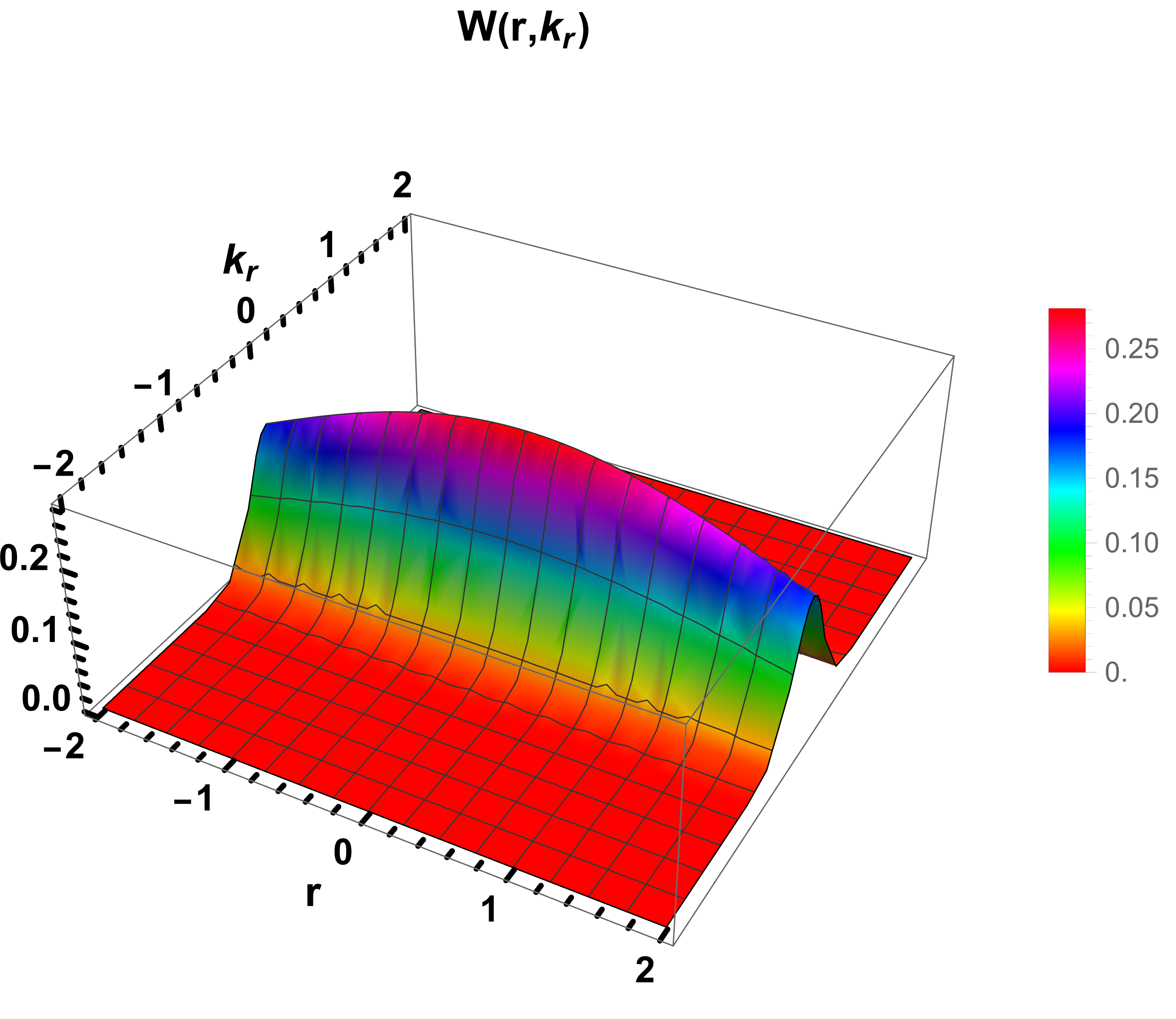}
  \caption{}
  \label{fig:WD1}
\end{subfigure}
\hfill
\begin{subfigure}{0.45\textwidth}
  \centering
  \includegraphics[width=\linewidth]{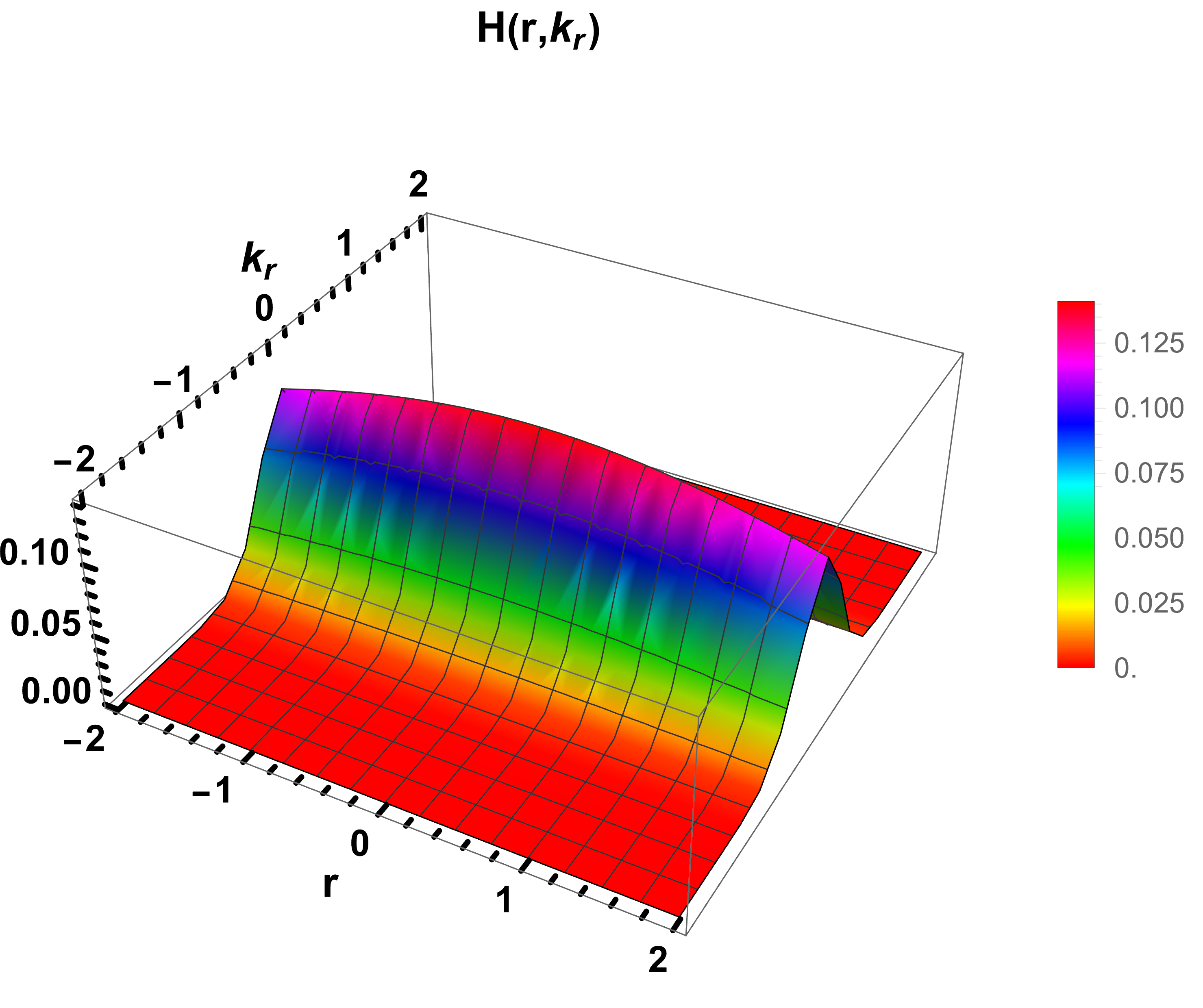}
  \caption{}
  \label{fig:WD2}
\end{subfigure}
\caption{Plot (~\ref{fig:WD1}) represents the Wigner distribution
$\mathcal{W}(r, k_{r})$ and plot (~\ref{fig:WD2}) represents the Husimi distribution $\mathcal{H}(r, k_{r})$. They are plotted for scattering length $(a)$ to be
$5\times 10^{-10}\,m$ and $N = 10$.}
\label{fig:wignerhusimi}
\end{figure}

Unlike our previous work ~\cite{ojha2025phase}, where the Wigner entropies were complex, the present ground state configuration yields strictly positive entropies. For the specific ground state wavefunction considered in this work — a $sech$-profile condensate in a harmonic trap with positive scattering length — the Husimi distribution is obtained from the Wigner distribution by Gaussian convolution with unit variance, and the resulting Wehrl entropy differs from the Wigner entropy by terms that are constant with respect to the scattering length a over the parameter range explored. This behavior is specific to the present configuration: it reflects the fact that the ground-state Wigner function remains positive definite and close to a Gaussian in phase space for the parameters considered, so that the Gaussian smoothing of the Husimi transform introduces only a fixed additive contribution to the entropy. This result does not hold in general — for states with significant Wigner negativity or strong non-Gaussian features, the Wehrl entropy can differ qualitatively from the Wigner entropy and provides genuinely distinct information about the quantum state~ ~\cite{wehrl1979relation,appleby1999generalized,rao2004cumulative,lieb1978proof}. We nevertheless present the Wehrl entropy for completeness, as it serves as a useful cross-check and as a reference for future studies in parameter regimes where Wigner negativity may arise. However, for the sake of completeness, we present a comparison between both distributions and their corresponding entropies. We begin by computing the Wigner and Husimi distributions for various scattering lengths, considering the particle number $N =10$. These distributions are then plotted as functions of the radial coordinate and the corresponding momentum along the radial direction. While both distributions are qualitatively similar in form, the numerical values obtained for each differ significantly (see Fig. ~\ref{fig:wignerhusimi}). We will focus on analyzing the Shannon entropies associated with both the Wigner and Husimi phase space distributions. While the Wehrl entropy is strictly real, the entropy derived from the Wigner distribution can, in general, be complex, in our case, it is not. In addition to the full distributions, we also investigate the entropies of their individual components. One might anticipate that the Shannon entropy of the Husimi distribution exceeds that of the Wigner distribution due to its coarse grained nature; however, a direct evaluation is essential, particularly when computing statistical correlations such as mutual information, which quantifies the relationship between the position $x$ and momentum $k$ variables. In addition, alternative measures, including Rényi entropies, are considered with the aim of defining real-valued Wigner entropies and correlation indicators.
\subsection{Marginals and Survival functions}\label{sec:becdistrib}
\begin{figure}[t]
\centering
\begin{subfigure}{0.45\textwidth}
  \centering
  \includegraphics[width=\linewidth]{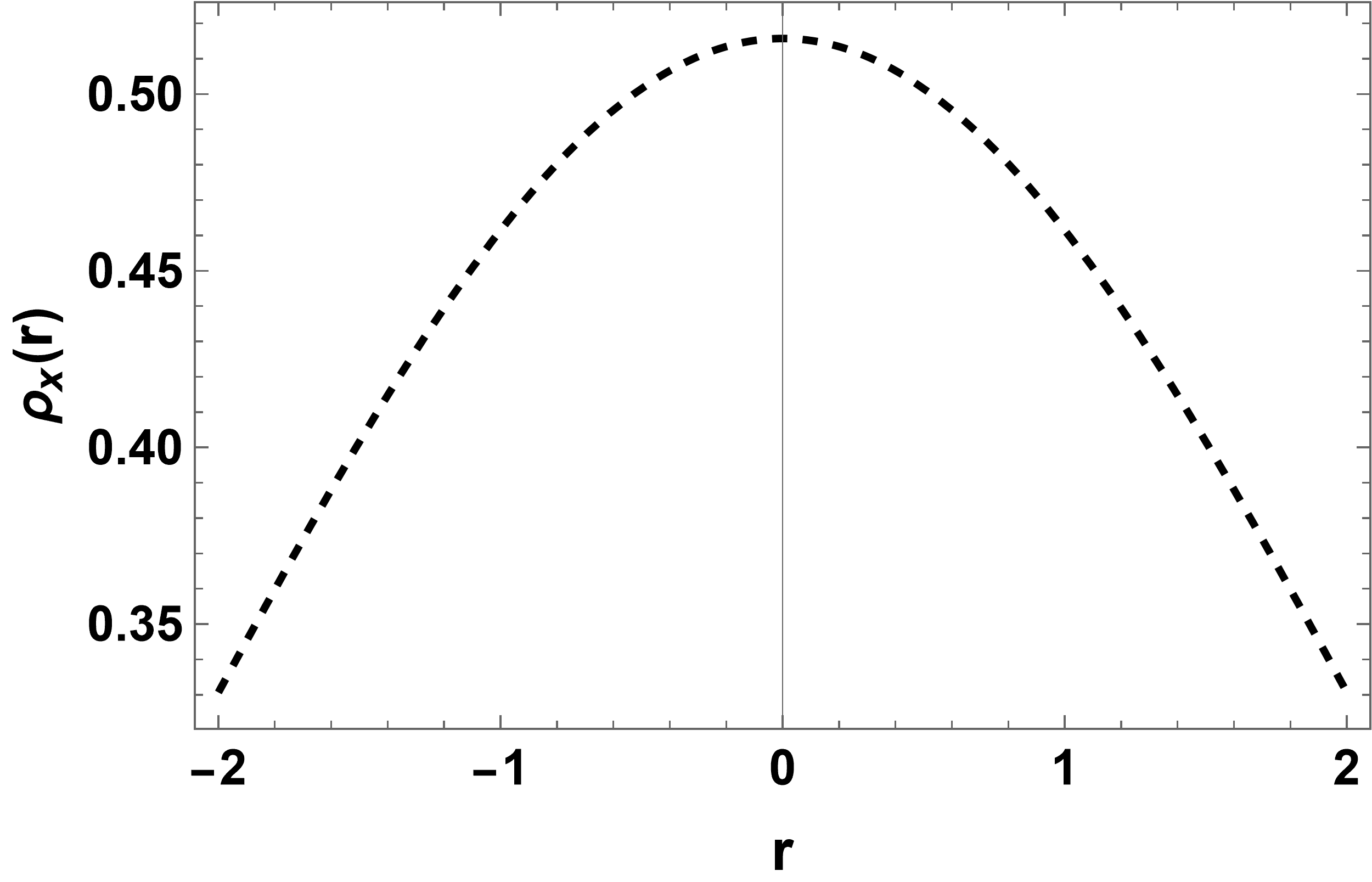}
  \caption{}
  \label{fig:WignerPos-marginals}
\end{subfigure}
\hfill
\begin{subfigure}{0.45\textwidth}
  \centering
  \includegraphics[width=\linewidth]{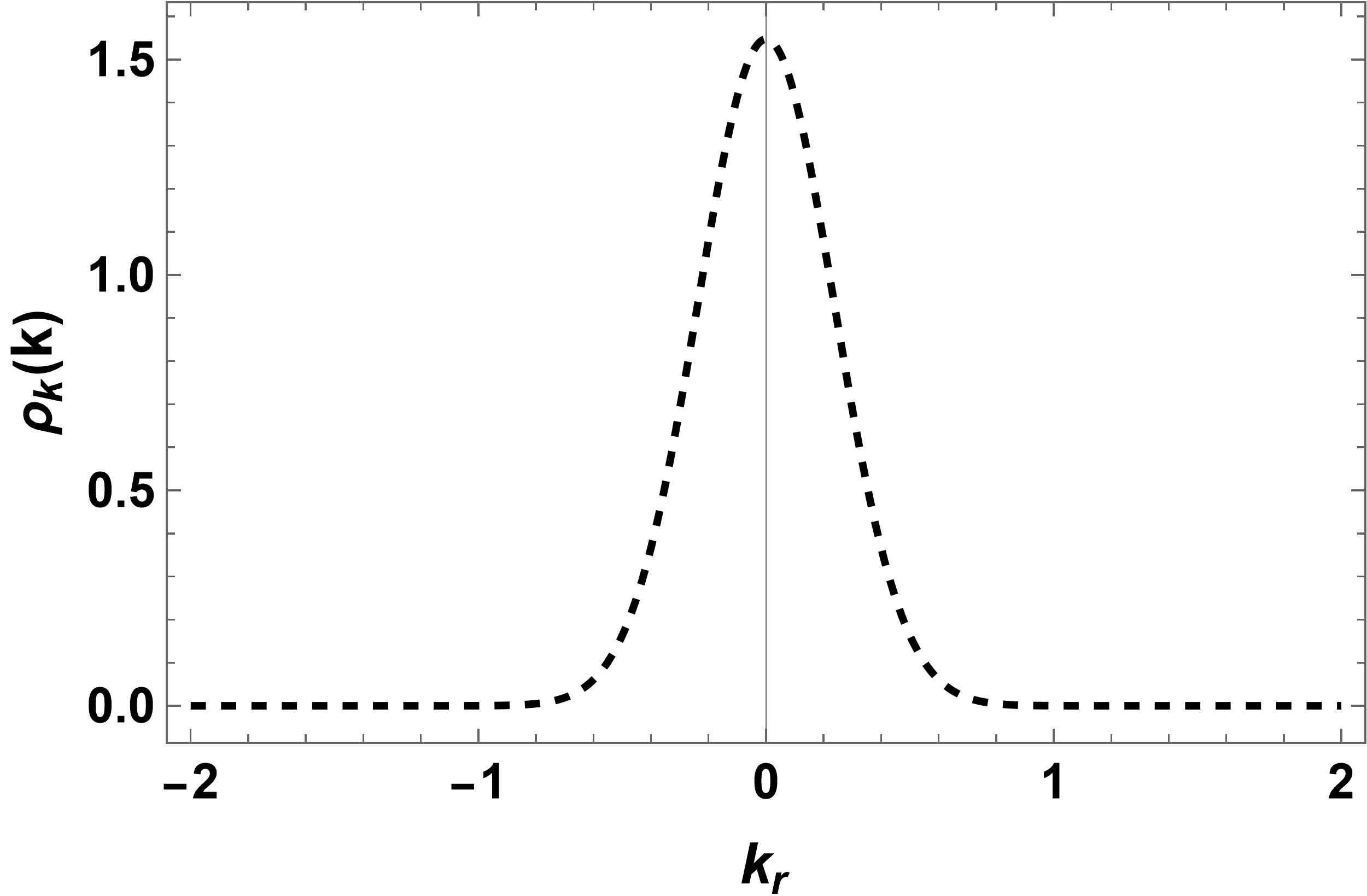}
  \caption{}
  \label{fig:WignerMom-marginals}
\end{subfigure}

\vspace{0.5cm}
\begin{subfigure}{0.45\textwidth}
  \centering
  \includegraphics[width=\linewidth]{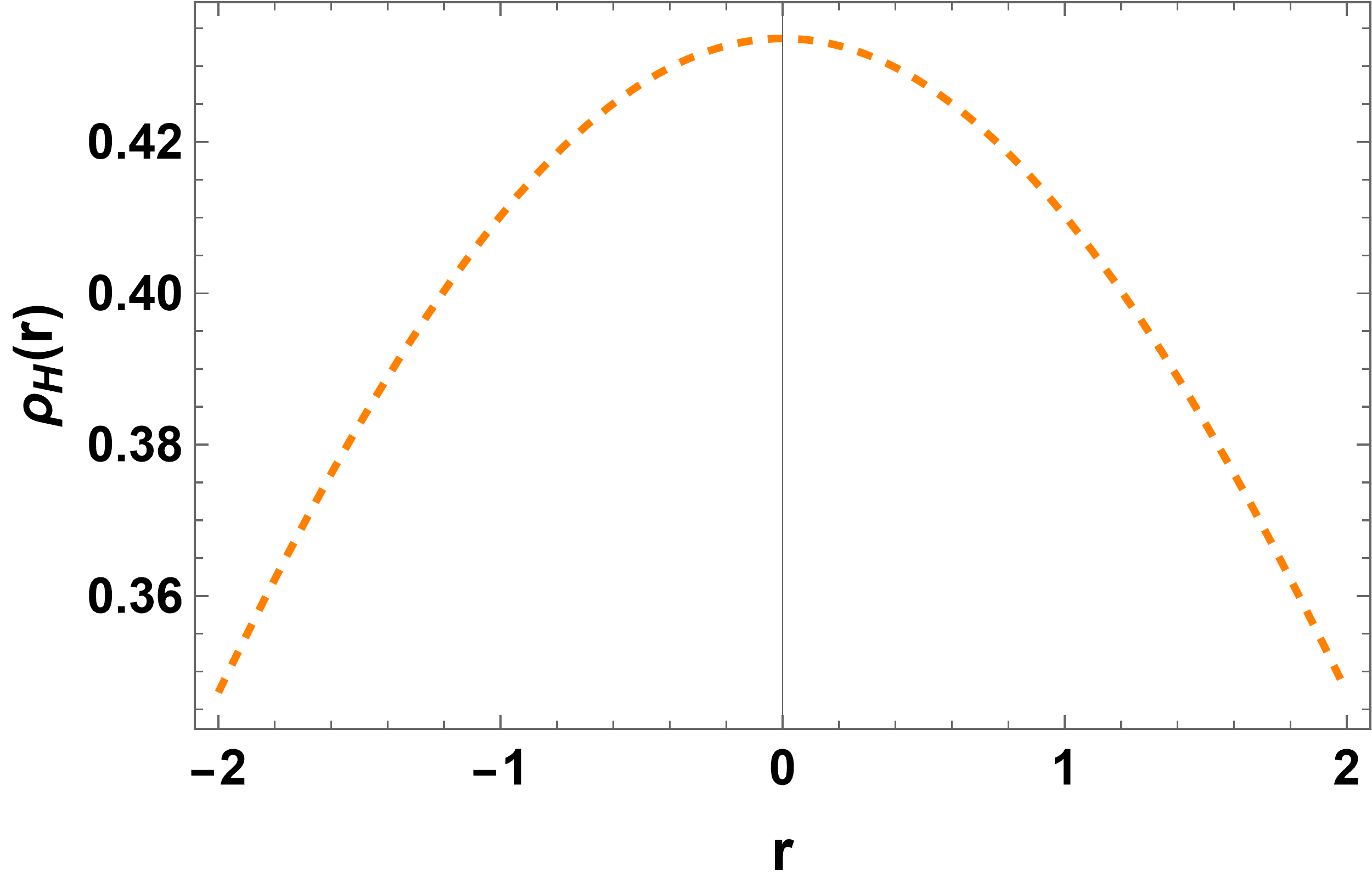}
  \caption{}
  \label{fig:HusimiPos-marginals}
\end{subfigure}
\hfill
\begin{subfigure}{0.45\textwidth}
  \centering
  \includegraphics[width=\linewidth]{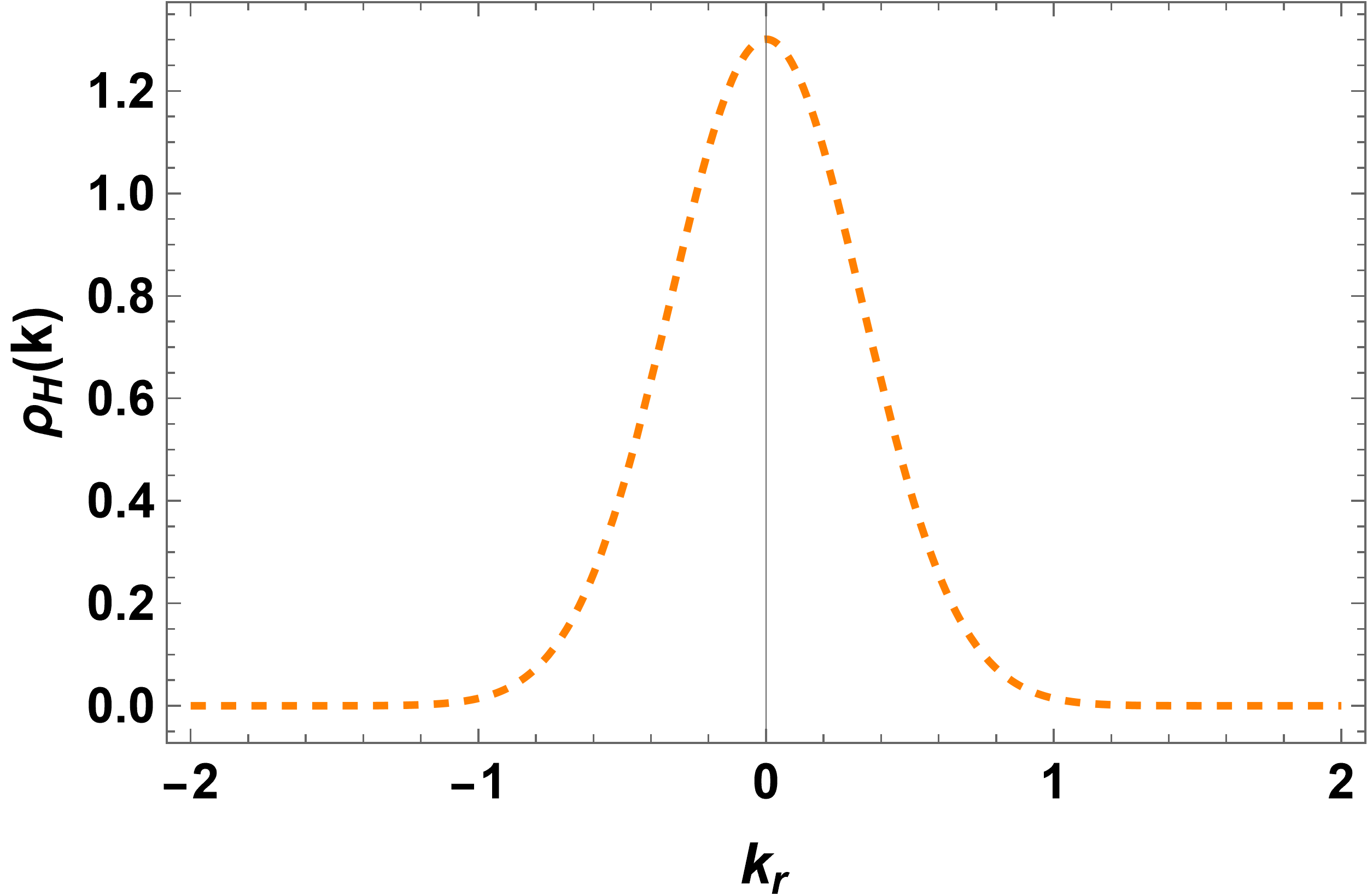}
  \caption{}
  \label{fig:HusimiMom-marginals}
\end{subfigure}

\caption{Plots (~\ref{fig:WignerPos-marginals}), (~\ref{fig:WignerMom-marginals}) represent the Wigner    distribution marginals in both position space and momentum space, respectively.  Plots  (~\ref{fig:HusimiPos-marginals}), (~\ref{fig:HusimiMom-marginals}) represent the Husimi distribution marginals in both position space and momentum space. They are plotted for scattering length $(a)$ to be $5\times 10^{-10}\,m$ and $N = 10$.  }
\label{fig:marginals}
\end{figure}

\begin{figure}[t]
\centering
\begin{subfigure}{0.45\textwidth}
  \centering
  \includegraphics[width=\linewidth]{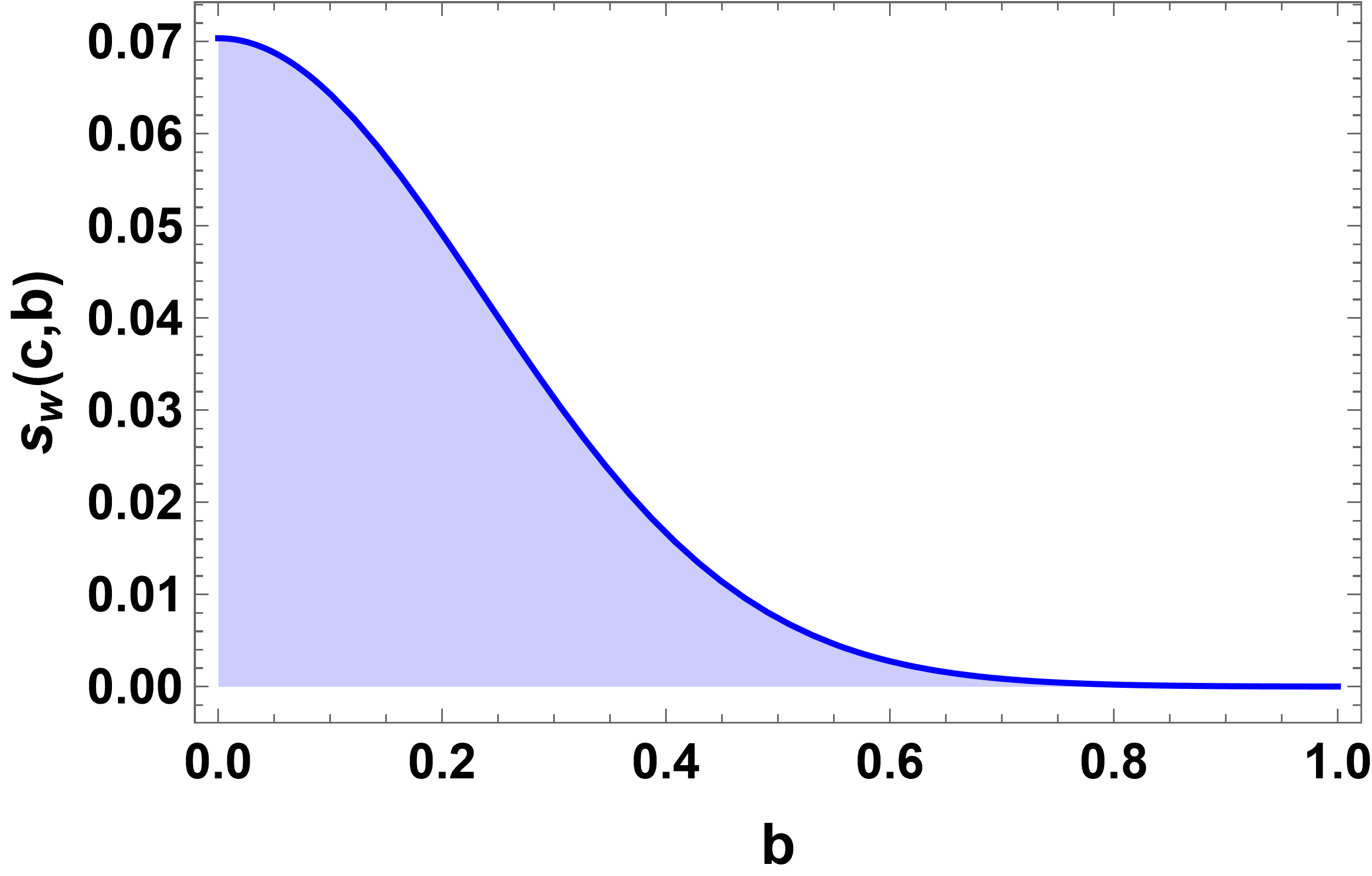}
  \caption{}
  \label{fig:3a}
\end{subfigure}
\hfill
\begin{subfigure}{0.45\textwidth}
  \centering
  \includegraphics[width=\linewidth]{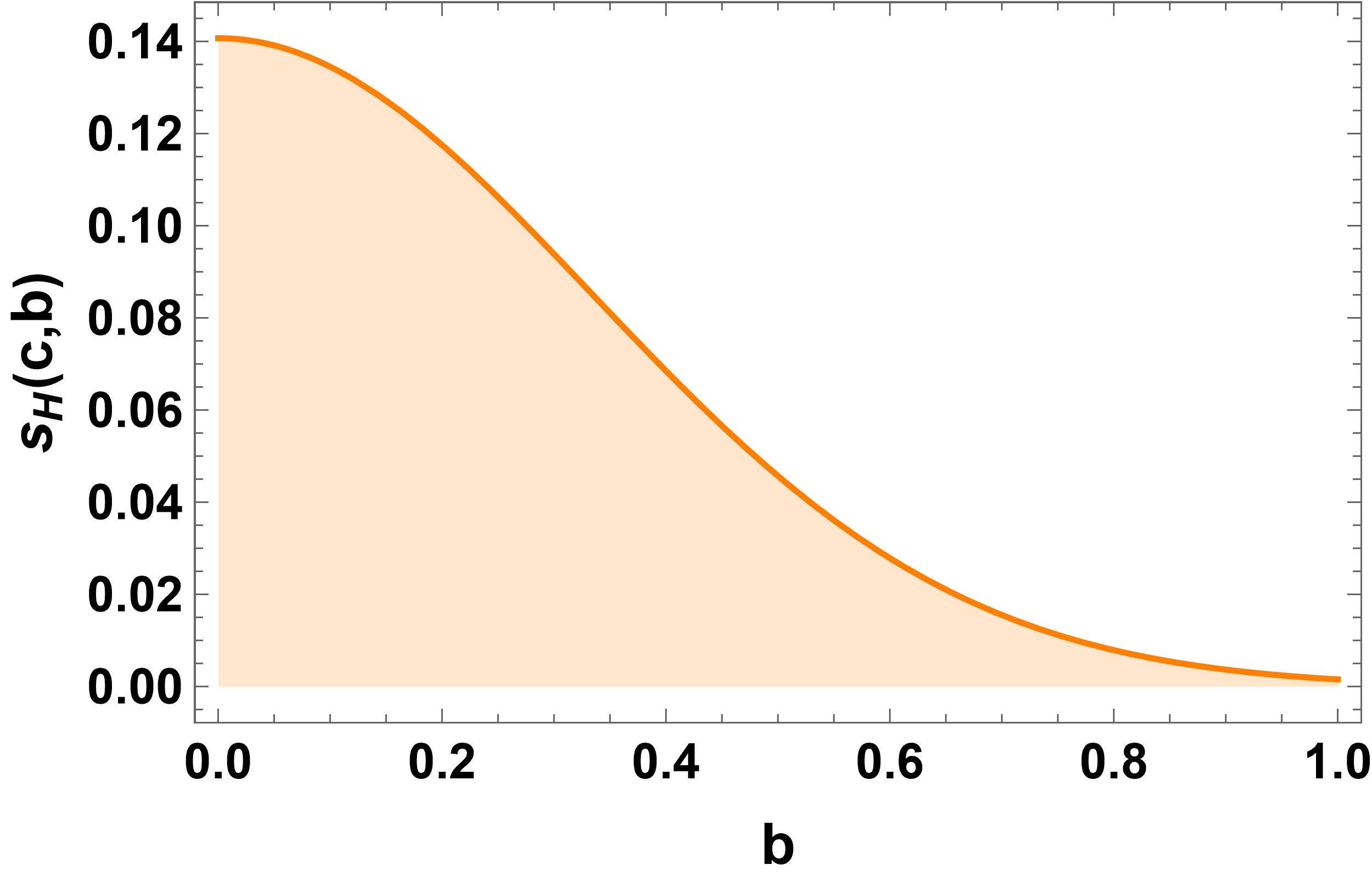}
  \caption{}
  \label{fig:3b}
\end{subfigure}
\caption{Plots~\ref{fig:3a} and~\ref{fig:3b} show the survival functions $s_{W}(c,b)$ and $s_{H}(c,b)$ obtained from the Wigner and Husimi distributions
respectively, plotted as functions of $b$ at fixed $c = 0$, for scattering length $a = 5\times 10^{-10}\,\mathrm{m}$ and $N = 10$. The survivals
are evaluated on the joint phase space $(r, k_{r})$, with one argument fixed and the other varied to expose the tail behaviour along the corresponding axis.}
\label{fig:wignerhusimisurvivals}
\end{figure}

The Wigner and Husimi distribution marginal densities can be compared in Fig.~\ref{fig:marginals}. The marginals in both position and momentum space are provided for both distributions. We can observe that a broad average is present in the Husimi marginal. Husimi marginals look broader because we have already averaged the state over a Gaussian phase space cell and so the broadening is just the variance of that kernel added to the original distribution.
After obtaining the phase space distributions, we compute the corresponding survival functions $s_{W}(c,b)$ and $s_{H}(a,b)$, shown in Fig. ~\ref{fig:wignerhusimisurvivals}, using the Wigner distribution, and the Husimi distribution, respectively.
The survival functions are plotted by fixing the value of $c = 0$ and varying the parameter $b$. Notably, for a positive definite distribution such as the Husimi distribution, the curves are monotonically decaying faster as a function of either $c$ or $b$ than that of Wigner distribution. Increasing $c$ shifts the surface uniformly downward (since the survival is monotonically non increasing in $c$) and progressively
truncates the contribution from the region of largest probability, but the qualitative ordering between the Wigner and Husimi survivals with the Husimi curve decaying more rapidly than the Wigner one is preserved across the range of $c$ values within the support of the distribution.

Physically, this demonstrates how coherent superpositions of quantum states occupy phase space in ways that are not captured by classical probability distributions alone. The survival functions also provide insight into the extent to which the quantum state is “spread out” or delocalized in phase space. From the plots, we observe that both the Wigner and Husimi distributions exhibit comparable delocalization, while the difference in the magnitude of the survival function indicates the amount of information lost. Since the Wigner distribution represents the true phase space distribution of the system, the loss of information is significantly smaller than in the Husimi distribution.

\begin{figure}[t]
\centering
\begin{subfigure}{0.45\textwidth}
  \centering
  \includegraphics[width=\linewidth]{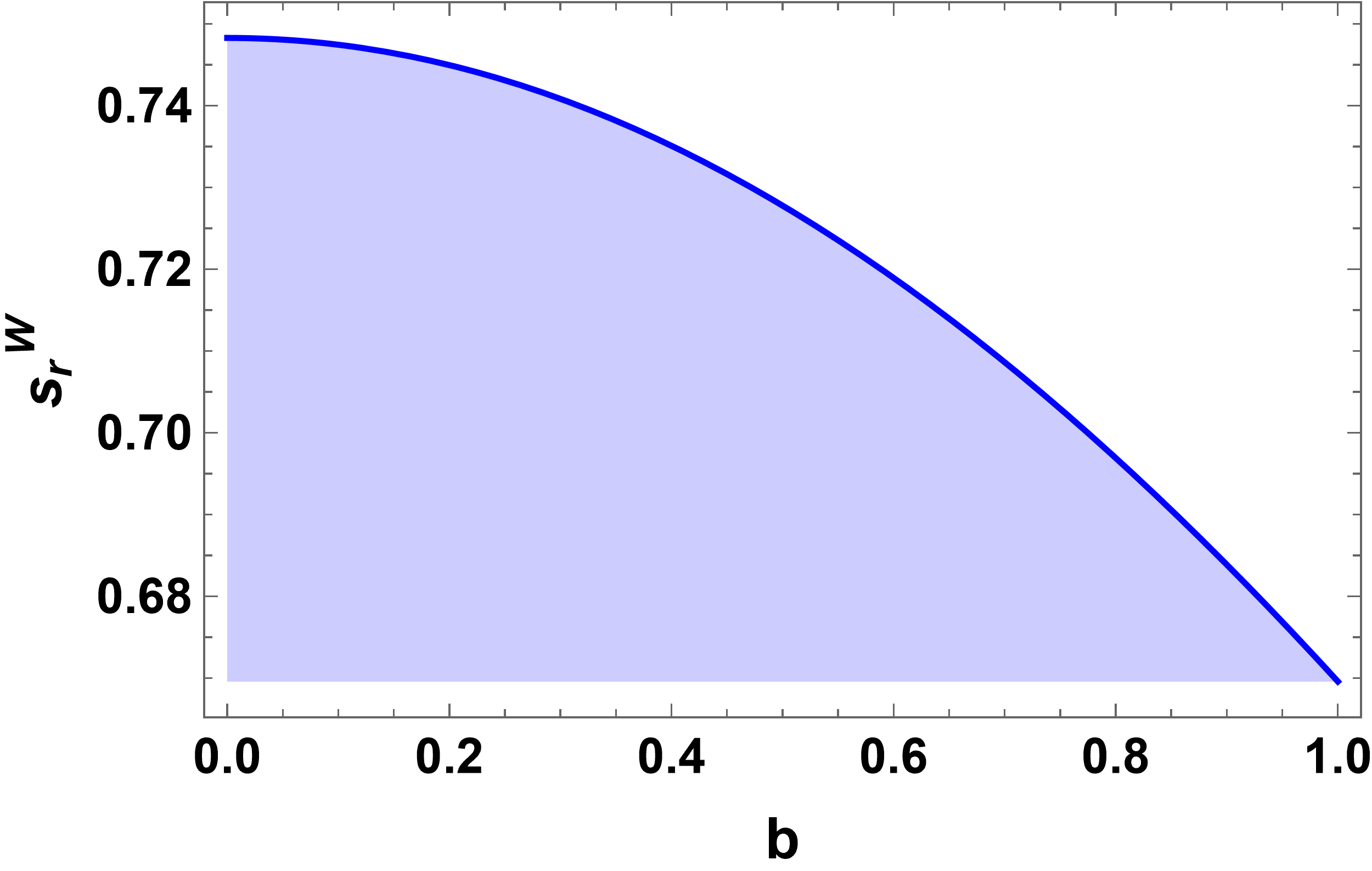}
  \caption{}
  \label{fig:image1259}
\end{subfigure}
\hfill
\begin{subfigure}{0.45\textwidth}
  \centering
  \includegraphics[width=\linewidth]{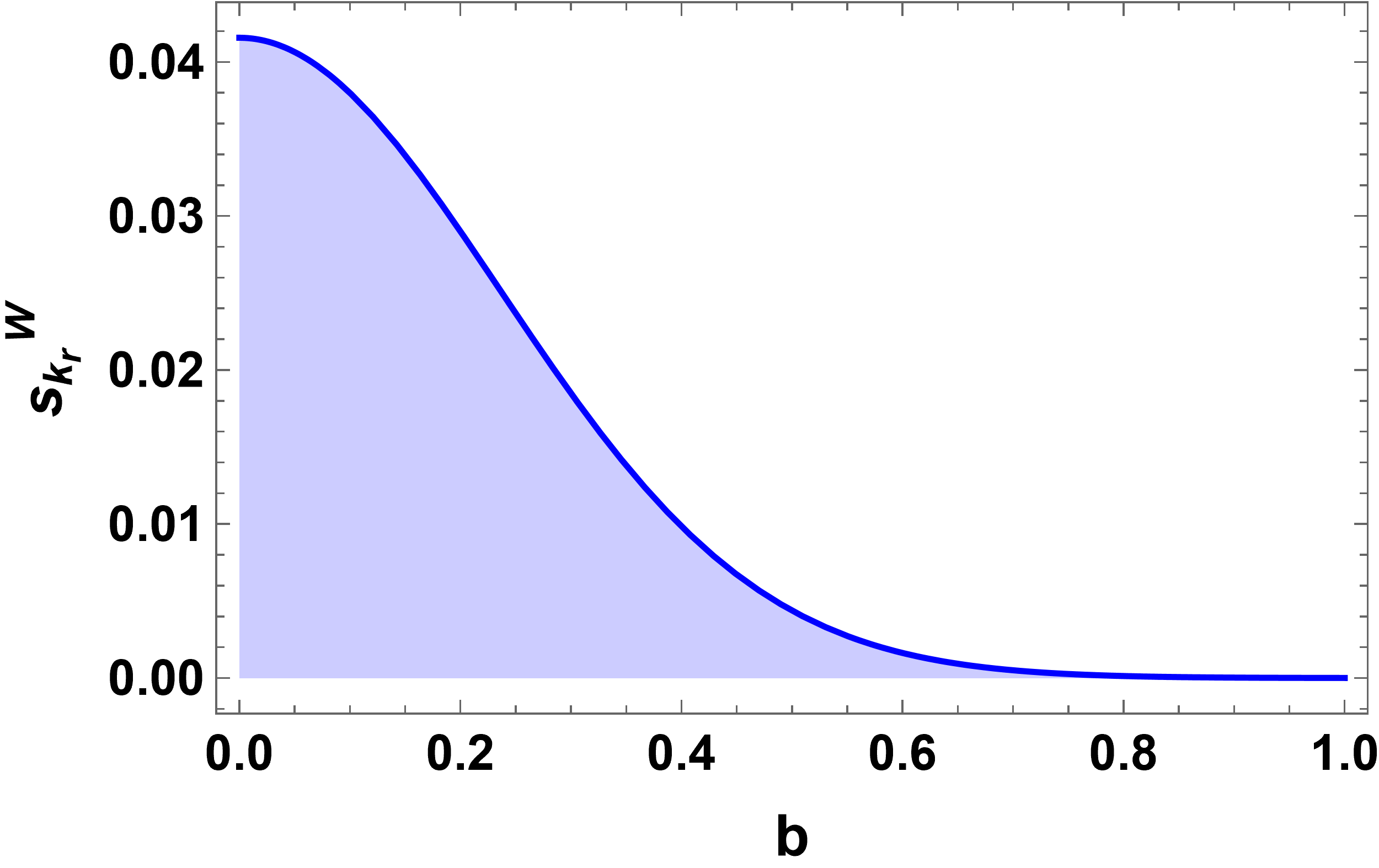}
  \caption{}
 \label{fig:image111}
\end{subfigure}

\vspace{0.5cm}

\begin{subfigure}{0.45\textwidth}
  \centering
  \includegraphics[width=\linewidth]{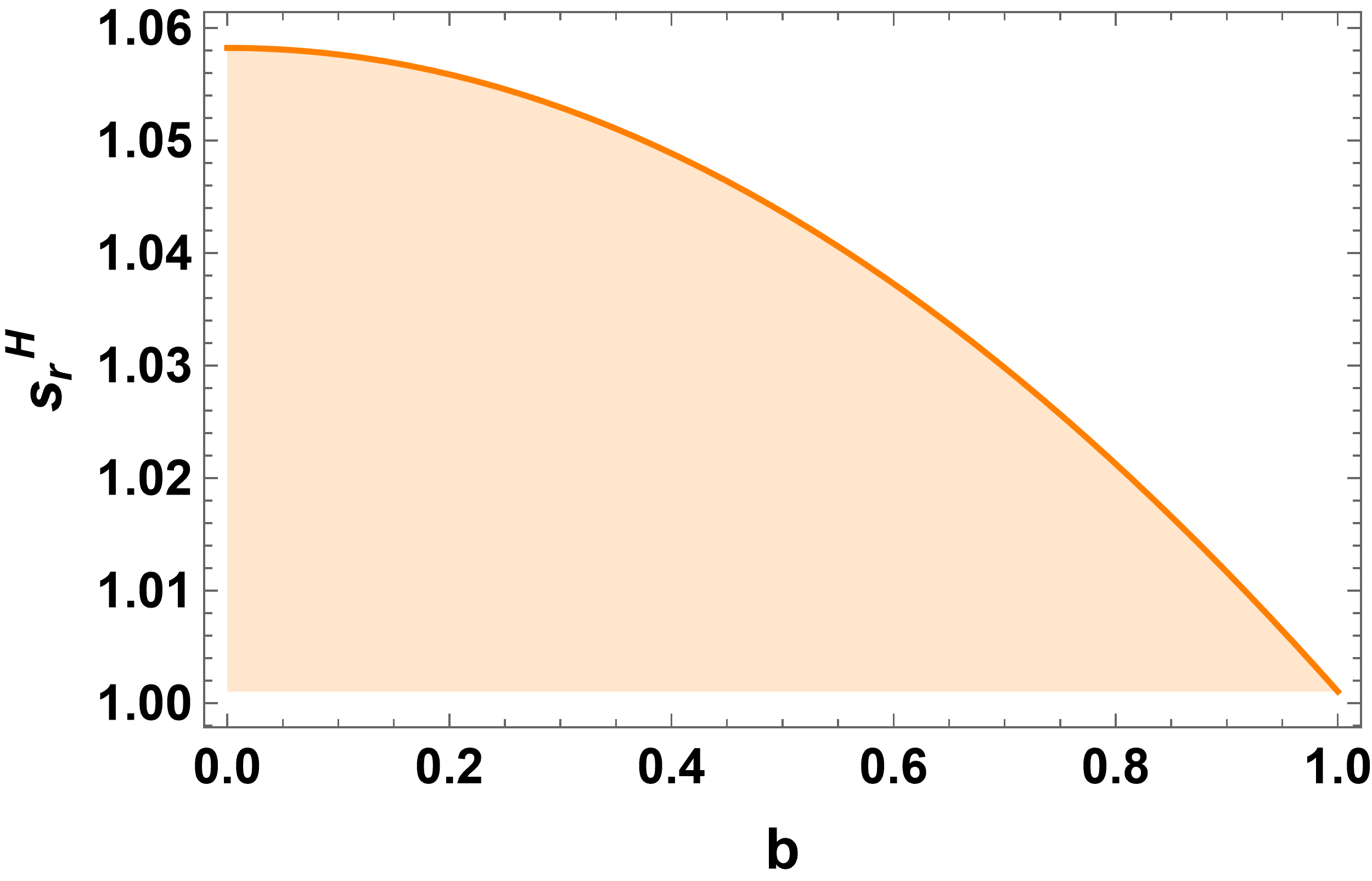}
  \caption{}
   \label{fig:image113}
\end{subfigure}
\hfill
\begin{subfigure}{0.45\textwidth}
  \centering
  \includegraphics[width=\linewidth]{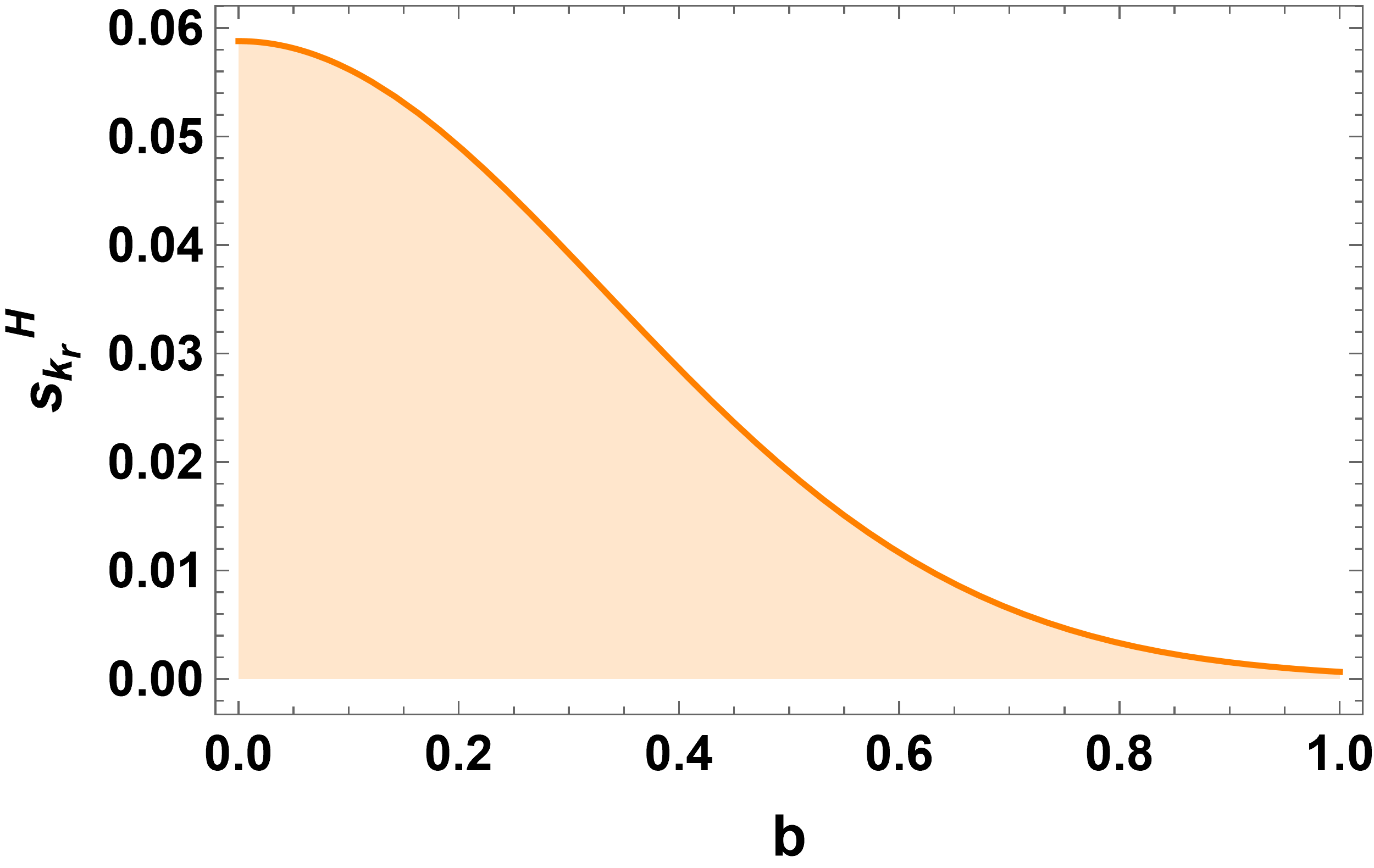}
  \caption{}
   \label{fig:image115}
\end{subfigure}
\caption{Plots (~\ref{fig:image1259}), (~\ref{fig:image111}) represent the survival functions from Wigner distribution marginals in position space and momentum space. Plots (~\ref{fig:image113}), (~\ref{fig:image115}) represent the survival functions from Husimi distribution marginals in position space and momentum space. They are plotted for scattering length $(a)$ to be $5\times 10^{-10}\,m$ for $N=10$. }
\label{fig:wignersurvivalsN10}
\end{figure}

Using the marginals obtained previously, we compute the survival functions constructed from both the Wigner and Husimi marginals, as shown in Fig.~\ref{fig:wignersurvivalsN10}. In contrast to the nodal structures observed in the anharmonic oscillator and harmonic oscillator cases~\cite{ojha2025phase,salazar2023phase}, no oscillatory wiggles appear here. Physically, this indicates a smoother phase space localization, suggesting the absence of strong quantum interference patterns that typically generate nodes in the survival probability. Instead, we observe only quantitative (numerical) deviations between the two distributions in both position and momentum space. In particular, the survival function derived from the Husimi marginal extends over a larger region of phase space in momentum space compared to the corresponding Wigner survival function, reflecting the intrinsically smoothed nature of the Husimi representation. The momentum space marginals exhibit qualitatively similar behavior, confirming that the observed trends reflect interaction-induced redistribution in the reduced probability densities rather than artifacts of a specific coordinate representation.   

\subsection{Entropies of the Wigner, Husimi distributions and their marginals}\label{sec:wigentro}
\begin{figure}[t]
\centering
\begin{subfigure}{0.45\textwidth}
  \centering
  \includegraphics[width=\linewidth]{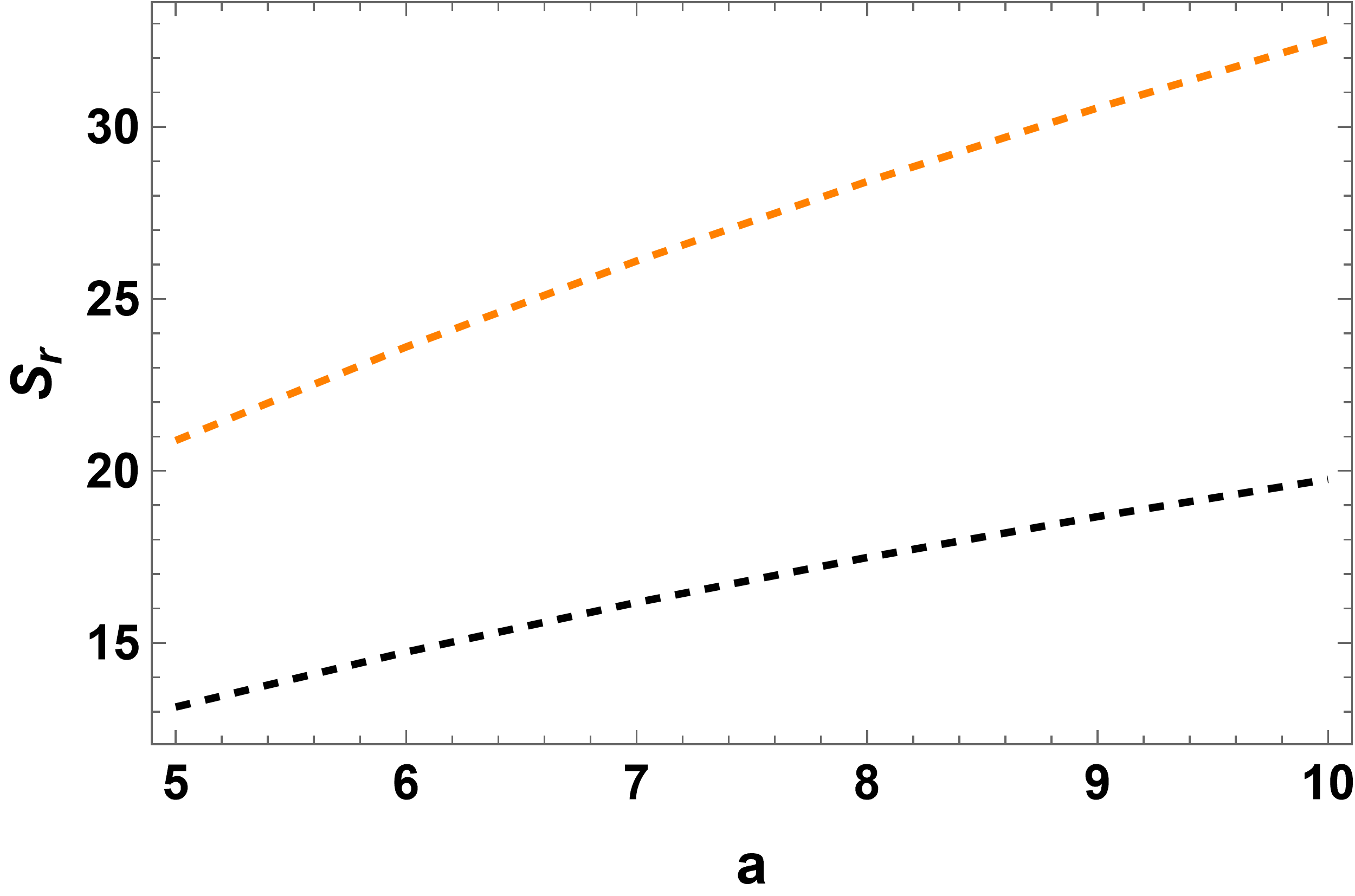}
  \caption{}
  \label{fig:image1955}
\end{subfigure}\hfill
\begin{subfigure}{0.45\textwidth}
  \centering
  \includegraphics[width=\linewidth]{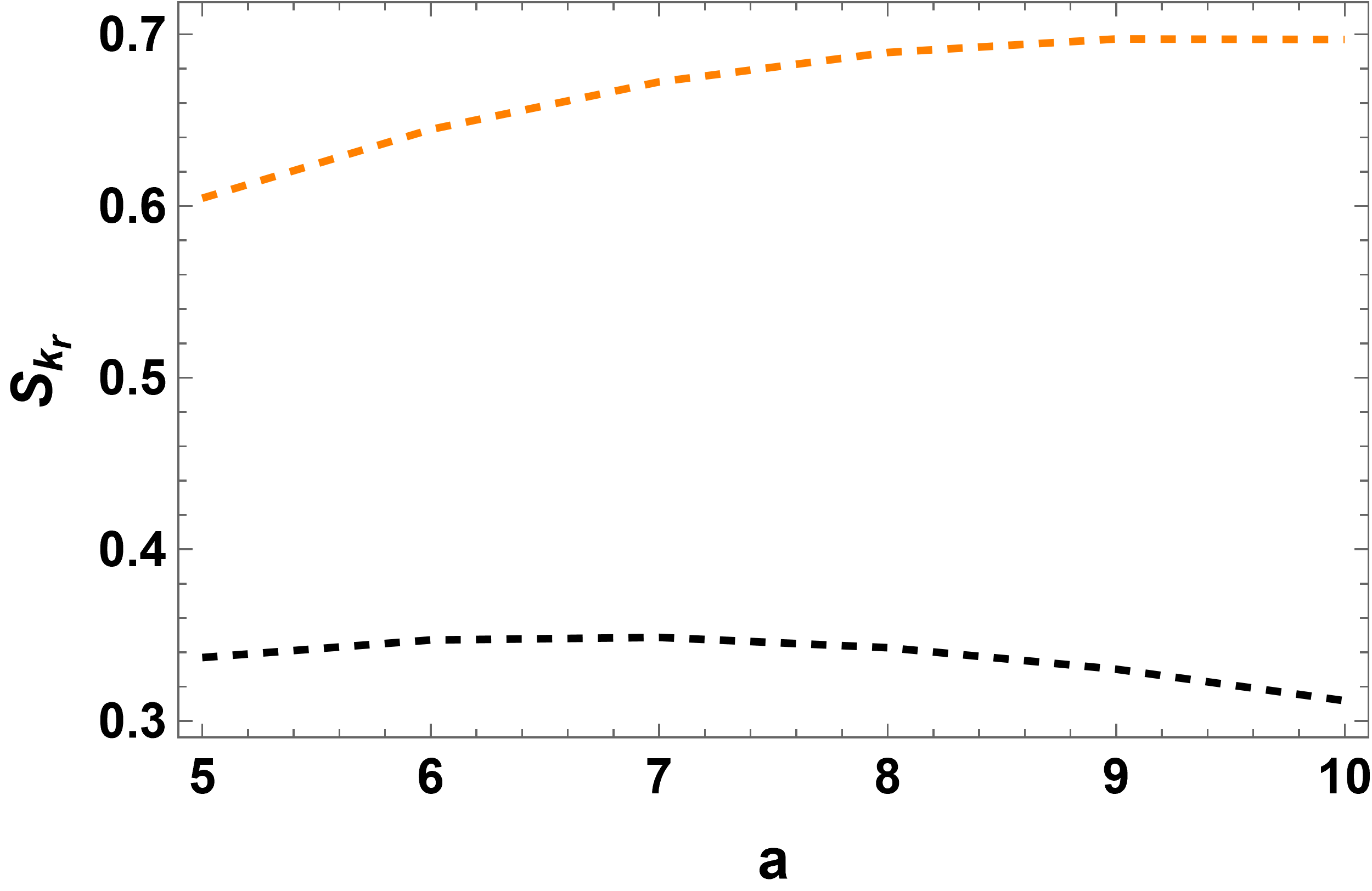}
  \caption{}
  \label{fig:image11055}
\end{subfigure}
\begin{subfigure}{0.45\textwidth}
  \centering
  \includegraphics[width=\linewidth]{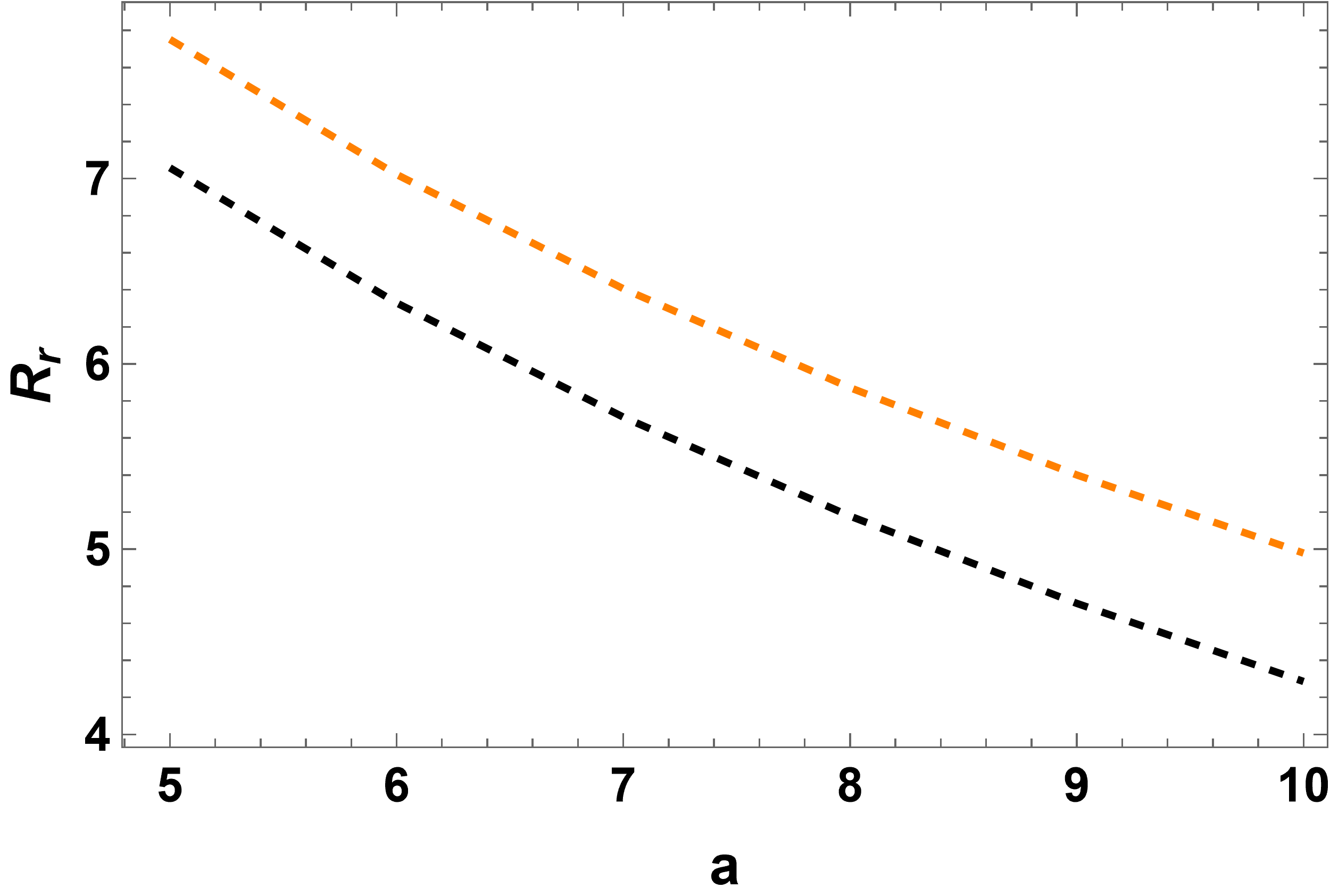}
  \caption{}
  \label{fig:i55}
\end{subfigure}\hfill
\begin{subfigure}{0.45\textwidth}
  \centering
  \includegraphics[width=\linewidth]{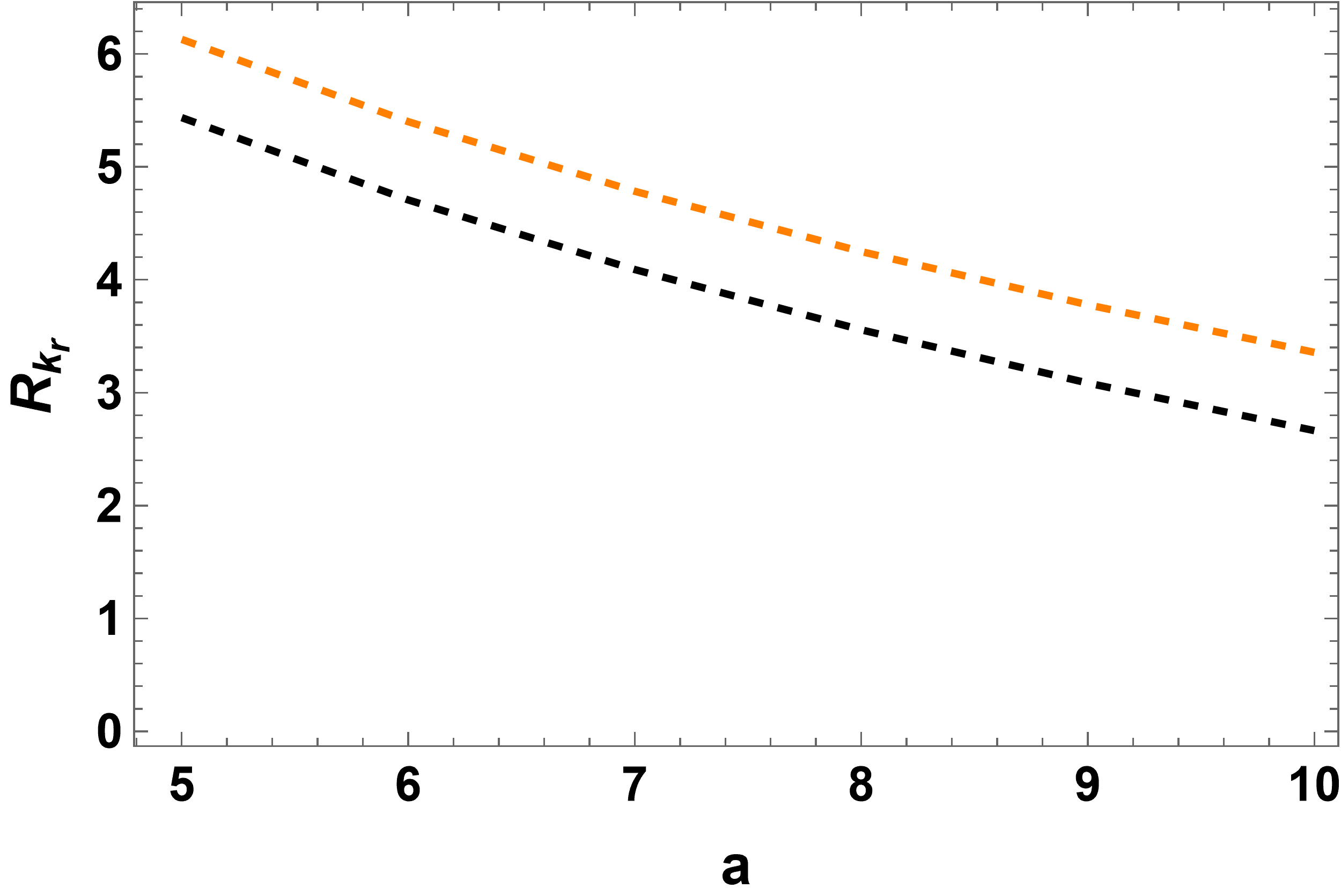}
  \caption{}
  \label{fig:i11055}
\end{subfigure}
\caption{Plots~(\ref{fig:image1955}) and~(\ref{fig:image11055}) show the Shannon entropy in position and momentum space, respectively, while plots~(\ref{fig:i55}) and~(\ref{fig:i11055}) present the Rényi entropy in position and momentum space, respectively. Entropies are obtained for different scattering lengths $(a\times 10^{-10}\, m)$. The Rényi entropies in plots~(\ref{fig:i55}) and~(\ref{fig:i11055}) are evaluated at $\alpha =4$}. Results obtained from the Wigner and Husimi marginals are shown in black and orange, respectively. Results obtained from the Wigner and Husimi distributions marginals are represented by black and orange lines, respectively.
\label{fig:ShannonentropyWH}
\end{figure}

We now present the discussion of entropies derived from the Wigner and Husimi marginals. Typically, the marginal entropies of the Wigner and Husimi functions are discussed before introducing the full distribution based entropies. This is done because all marginal distributions are positive definite, and there are no issues associated with complex valued entropies. Such an approach provides a clearer comparison between the different entropic quantities and allows a direct assessment of the differences between the Wigner and Husimi marginals. The montonically decrease in slope present in the survival functions (Figs.~(\ref{fig:image113}) and~(\ref{fig:image115})) indicates stronger localization (smaller entropy values) compared to the Husimi distribution. The loss of fine structure in the Husimi distribution relative to the Wigner distribution is responsible for the larger entropic values observed for the Husimi case. Figure~\ref{fig:ShannonentropyWH} shows the behavior of the Shannon and Rényi entropies in both position and momentum space for the Wigner and Husimi marginals, respectively, as a function of the scattering length $(a)$. The numerical value of the Shannon entropy increases with increasing scattering length. However, all entropies obtained from the Wigner marginals are smaller in magnitude than those derived from the Husimi for a given value of $a$. The Rényi entropies are shown in Fig.~(\ref{fig:i11055}). We also note that the Rényi entropy remains constant and does not depend on $a$ for $\alpha=2$, where $\alpha$ is the parameter appearing in Eq.~(\eqref{eq:RENyipos}). This contrasts with the behavior of the Rényi entropy plotted for $\alpha=4$. Unlike the Shannon entropy, it decreases with increasing scattering length $(a)$. The Rényi entropy obtained from the Wigner distribution marginals is smaller than that from the Husimi distribution. The same physical reasoning discussed for the Shannon entropy applies to the Rényi entropy as well. 

\begin{figure}[H]
    \centering
    \begin{subfigure}[t]{0.45\textwidth}
        \centering
        \includegraphics[width=\textwidth]{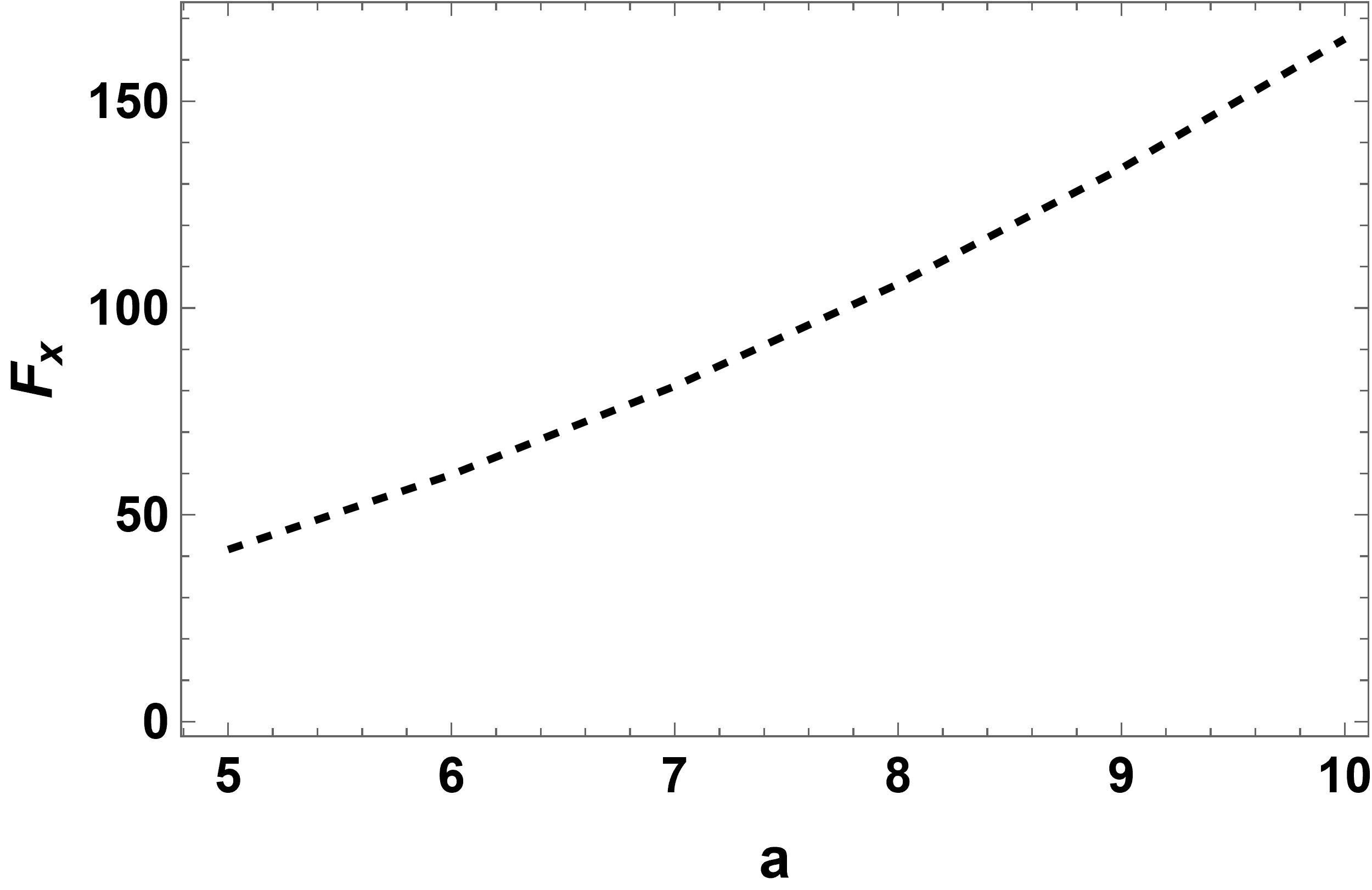}
        \caption{Position space}
         \label{fig:image19}
    \end{subfigure}
\hfill
     \vspace{1em}
    \begin{subfigure}[t]{0.45\textwidth}
        \centering
        \includegraphics[width=\textwidth]{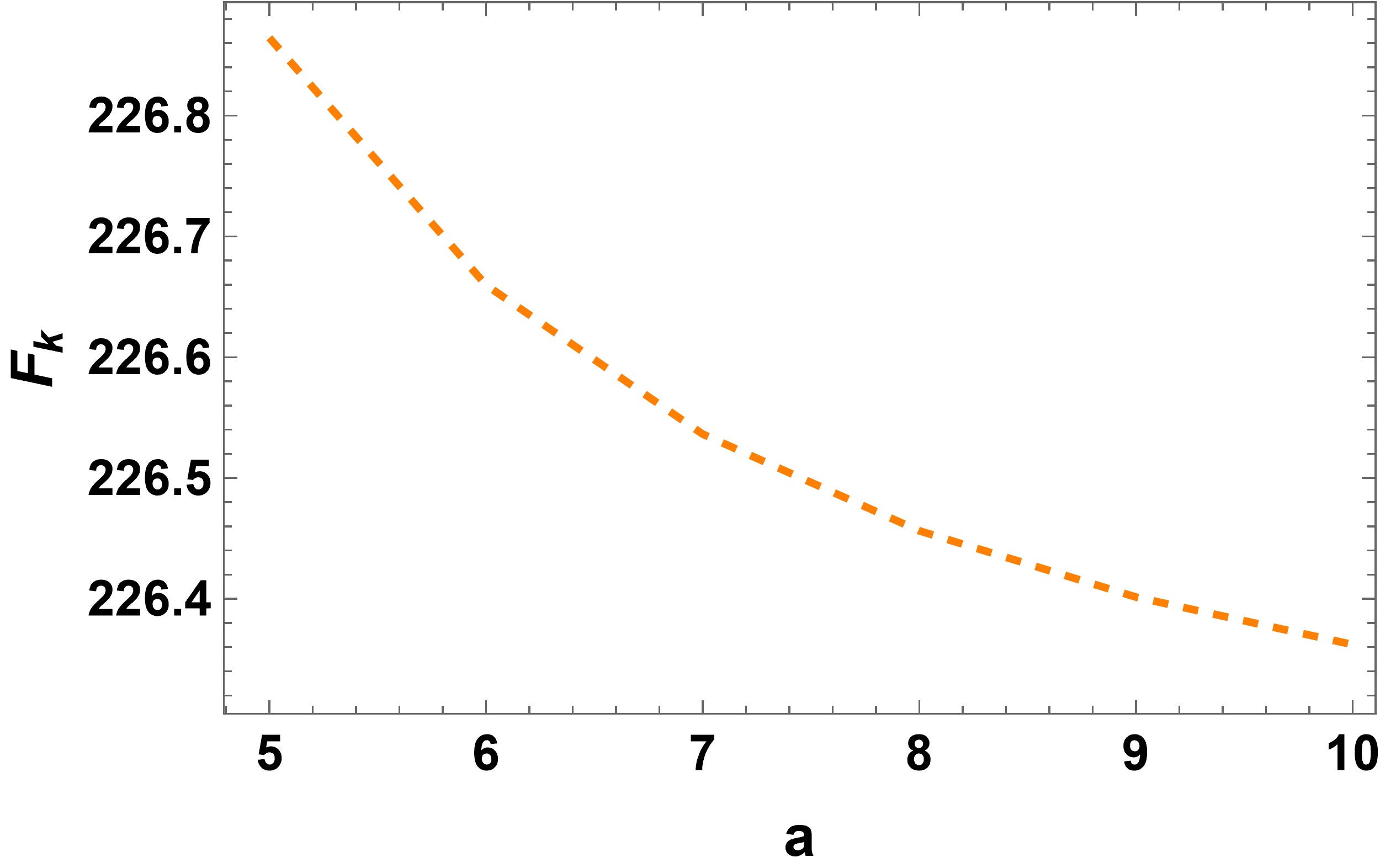}
        \caption{Momentum space}
         \label{fig:image20}
    \end{subfigure}
    \caption{Variation of the Fisher information measure $\mathcal{F}$ with the scattering length $a \times 10^{-10}\,\mathrm{m}$ at $N=10$. Plot~(a) shows the position-space Fisher information $F_{r}$ and plot~(b) shows the momentum-space Fisher information $F_{k_{r}}$. The product $F_{r}\cdot F_{k_{r}} \geq 4$ is satisfied throughout the parameter range.}
\label{fig:Fisherplot}
\end{figure}

The observed entropy trends can be understood within the framework of quantum information theory, where Shannon and related entropic measures quantify the information content and the degree of delocalization of quantum states in phase space. Increasing $N$ generally enhances configurational complexity, which leads to larger information entropy and a more effectively mixed description of the state~\cite{wehrl1978general}. The results are also consistent with entropic uncertainty relations, which imply that the total information content of the system remains bounded~\cite{bialynicki2006formulation}. Overall, these findings indicate that entropy, viewed as an information theoretic measure, provides a sensitive probe of the interplay between interaction strength, spatial delocalization, and quantum coherence in many body systems. After computing the entropies from the Wigner and Husimi marginals, we next evaluate the entropies derived from the Husimi distribution, commonly referred to as the Wehrl entropy. Although the Wigner distribution can, in general, exhibit negative regions, in the present ground state configuration considered here it remains positive definite over the entire phase space. Consequently, the Shannon entropy derived from the Wigner distribution is well defined and real valued in our case. We nevertheless include the Wehrl entropy for comparison (see Fig.~\ref{fig:7a}), since the Husimi distribution provides a coarse grained phase space representation that highlights semiclassical features. We observe that it increases with increasing scattering length. Physically, the growth of the Wehrl entropy with scattering length reflects the interaction induced spreading of the Husimi distribution, which enlarges the effective phase space volume occupied by the quantum state. In Fig.~\ref{fig:Fisherplot}, we present the Fisher information obtained from the marginals of the Wigner distribution. With increasing scattering length, the Fisher information increases in position space and decreases in momentum space. Consequently, their product satisfies the uncertainty relation formulated in terms of Fisher information \cite{fisher1}. 
The monotonic increase of the position-space marginal Shannon entropy $S^{W}_{r}$ of the Wigner distribution [black curve in Fig.~\ref{fig:image1955}] and of the Wehrl entropy $S_{H}$ of the Husimi distribution [Fig.~\ref{fig:7a}] with the scattering length $a$ reflects enhanced spatial delocalization caused by repulsive interactions, which broaden the condensate wavefunction and reduce positional localization. In the present configuration the Wigner function is positive definite and the full phase-space Shannon entropy $S_{W}$ is therefore well-defined; we find $S_{W} < S_{H}$ across the parameter range, with the offset reflecting the fixed additive contribution of the Gaussian convolution kernel intrinsic to the
Husimi transform. Both $S_{W}$ and $S_{H}$ are now overlaid in Fig.~\ref{fig:7a} as a cross-check of this relationship. We emphasize that Fisher in formation and Shannon / Wehrl entropy are not redundant measures of localisation. The Shannon and Wehrl
entropies are functionals of the density that quantify its support --- how much of phase space distribution effectively occupies. The Fisher information, in contrast, is a functional of the logarithmic gradient of the density and quantifies its local sharpness. A density that broadens overall while developing sharper local features can simultaneously produce a larger
Shannon entropy and a larger Fisher information without
contradiction; the two measures are sensitive to complementary aspects of the same distribution. In the present $sech$ - profile ground state, increasing $ a $ broadens the position - space envelope, raising
the Shannon entropy, while at the same time sharpening the relative spatial modulation of the density ( the $ a $ - dependent argument of the $sech$ grows with $ a $), raising $ F _{ r }$. The complementary decrease of
$ F _{ k _{ r }}$ reflects the smoothing of the momentum - space profile imposed by Fourier reciprocity throughout, and the product $ F _{ r } \cdot F _{ k _{ r }} \geq 4$ is satisfied throughout the parameter range.

The behavior of Fisher information requires careful interpretation. In position space, the Fisher information $\mathcal{F}_{r}$ increases with increasing scattering length, while in momentum space $\mathcal{F}_{k}$ decreases, and their product satisfies the uncertainty bound $\mathcal{F}_{r}\cdot \mathcal{F}_{k} \geq 4$ throughout. This behavior might appear to contradict the notion that repulsive interactions broaden and smooth the condensate wavefunction, since smoothing generically reduces local gradients and would be expected to decrease Fisher information. The resolution lies in distinguishing between two competing effects of the repulsive interaction. On the one hand, the condensate broadens in real space, reducing the overall amplitude of the wavefunction. On the other hand, repulsive interactions generate a spatially varying phase across the condensate profile, and the Fisher information is sensitive to the logarithmic derivative of the probability density, which can increase even as the distribution broadens if the interaction introduces sharper relative variations in the density profile. In the present sech-profile ground state, the interaction-dependent prefactor in the wavefunction introduces such variations: as $a$ increases, the spatial modulation of the density becomes more pronounced relative to the overall envelope, driving $\mathcal{F}_{r}$ upward. The complementary decrease of $\mathcal{F}_{k}$ reflects the reciprocal relationship imposed by the uncertainty principle — as real space structure becomes more pronounced, momentum space structure becomes smoother. This complementary behavior is consistent with the localization-delocalization tradeoff discussed by Dong and Sun ~\cite{sun2013quantum,shi2018quantum} and its analog in the present interaction-driven context.

The entropic measures employed in this work are constrained by well-known entropic uncertainty relations ~\cite{robertson1929uncertainty,kennard1927quantenmechanik,sen2011statistical}. For single particle systems, Shannon and Rényi entropies of position and momentum distributions obey bounds such as the Białynicki-Birula–Mycielski (BBM) inequality ~\cite{bialynicki1975uncertainty,coles2017entropic}, which describes entropy-based localization, including those reported by Sun et al. In the present work, although the system is intrinsically many-body and nonlinear, the phase space description constructed from the Gross-Pitaevskii wavefunction ~\cite{pitaevskii1961vortex,gross1961structure} continues to satisfy analogous entropic uncertainty relations at the level of one-body marginals. Importantly, interactions do not violate these bounds but instead control how uncertainty is redistributed between conjugate variables. In the weak interaction or low particle number regime, our results smoothly reduce to established single particle behavior, recovering standard entropy localization correspondences. In contrast to the potential-driven localization mechanisms analyzed in the Sun et al ~\cite{solaimani2020quantum,carrillo2022shannon,gil2022quantum,santana2023quantum,carrillo2021shannon,shi2018quantum,torres2018radial,valencia2015quantum,sun2013quantum1,sun2013quantum,serrano2016information,sun2013quantum} framework, where entropy variations are governed by external confinement parameters, the trends observed here are governed by interaction-induced restructuring of phase space distributions. As interaction strength and particle number increase, enhanced many-body correlations and nonlinear phase space structure drive delocalization without saturating single particle uncertainty bounds. This demonstrates that the observed entropy trends are not merely a reformulation of known single-particle results in a different representation, but arise from interaction-driven changes in phase space uncertainty in a many-body system.
\subsection{Quantifying Information Differences: KL, Jeffreys, and Rényi divergences}
\begin{figure}[H]
\centering
\begin{subfigure}{0.40\textwidth}
        \centering
        \includegraphics[width=\linewidth]{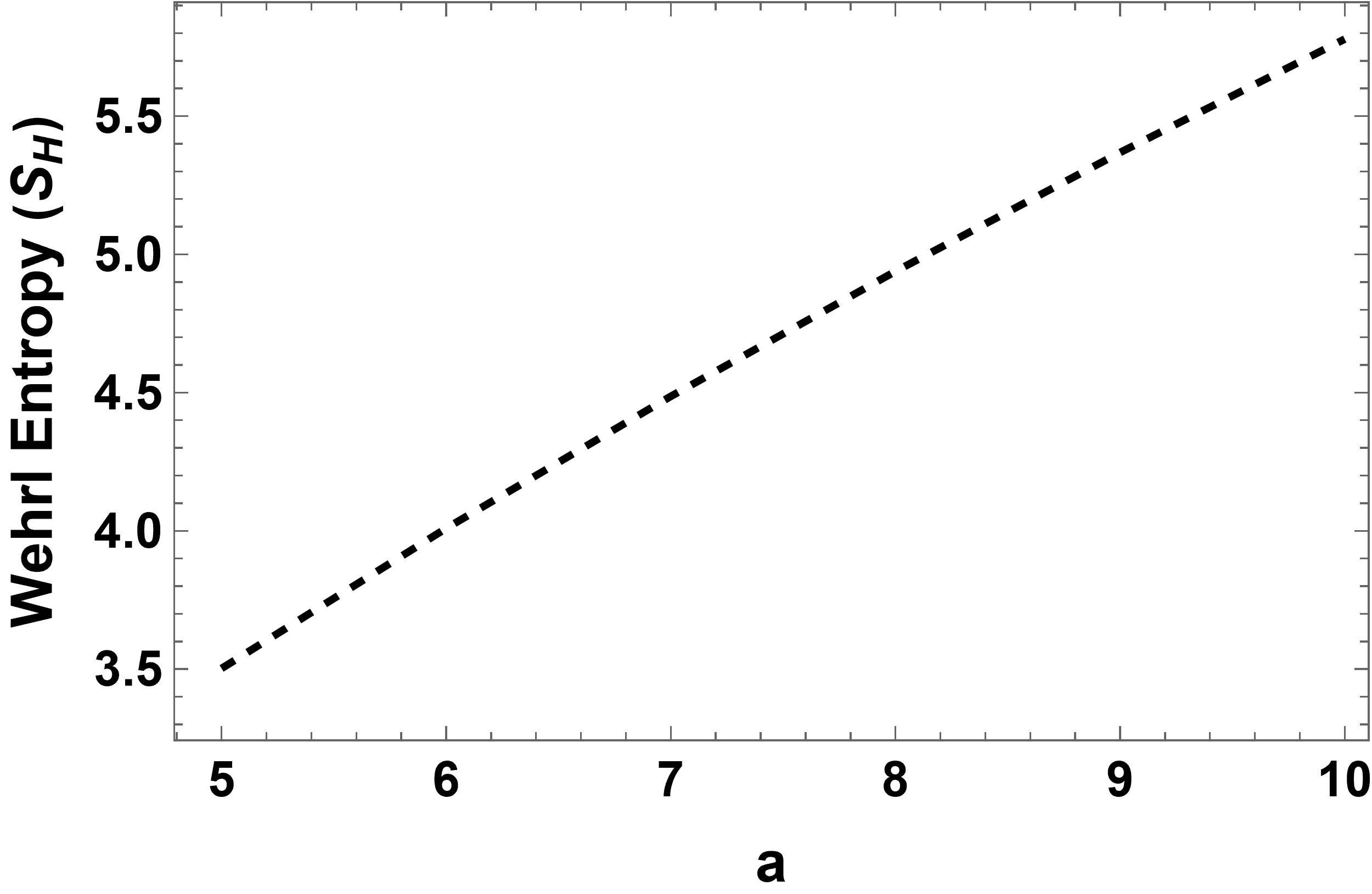} 
        \subcaption{}
         \label{fig:7a}
    \end{subfigure}
\begin{subfigure}{0.40\textwidth}
  \centering
  \includegraphics[width=\linewidth]{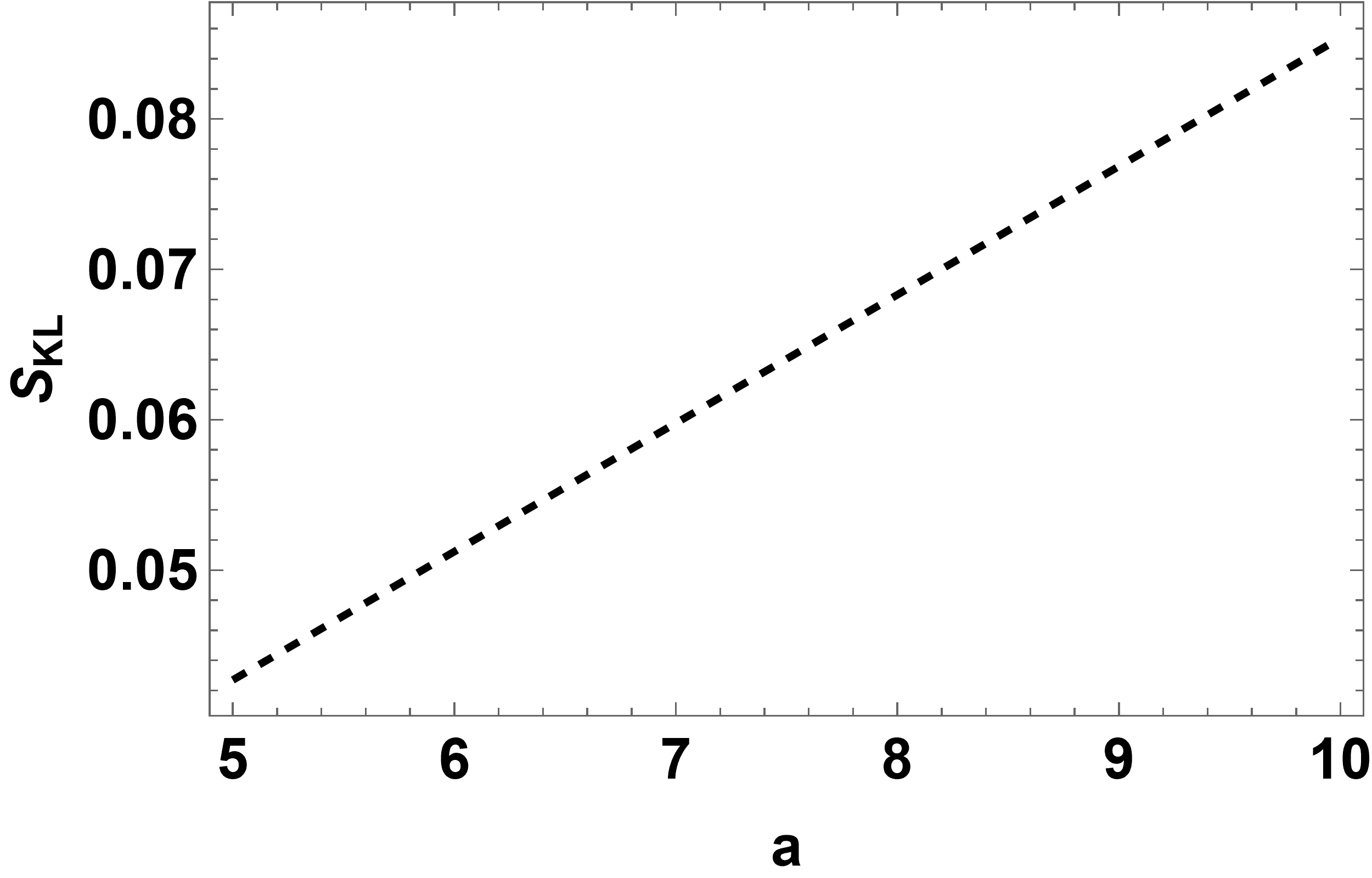}
  \caption{}
  \label{fig:7b}
\end{subfigure}
\begin{subfigure}{0.40\textwidth}
  \centering
  \includegraphics[width=\linewidth]{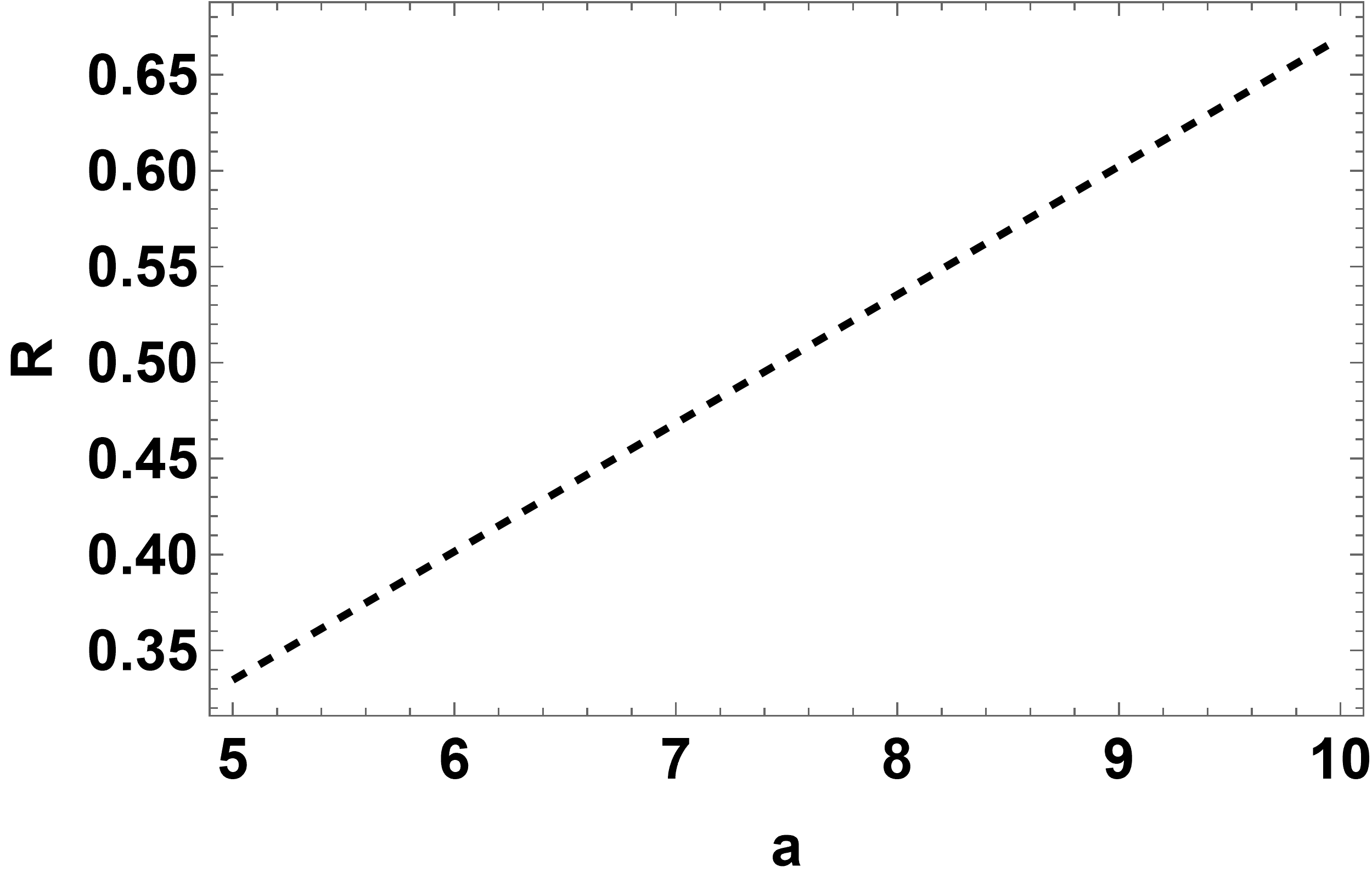}
  \caption{}
  \label{fig:7c}
\end{subfigure}
\begin{subfigure}{0.40\textwidth}
  \centering
  \includegraphics[width=\linewidth]{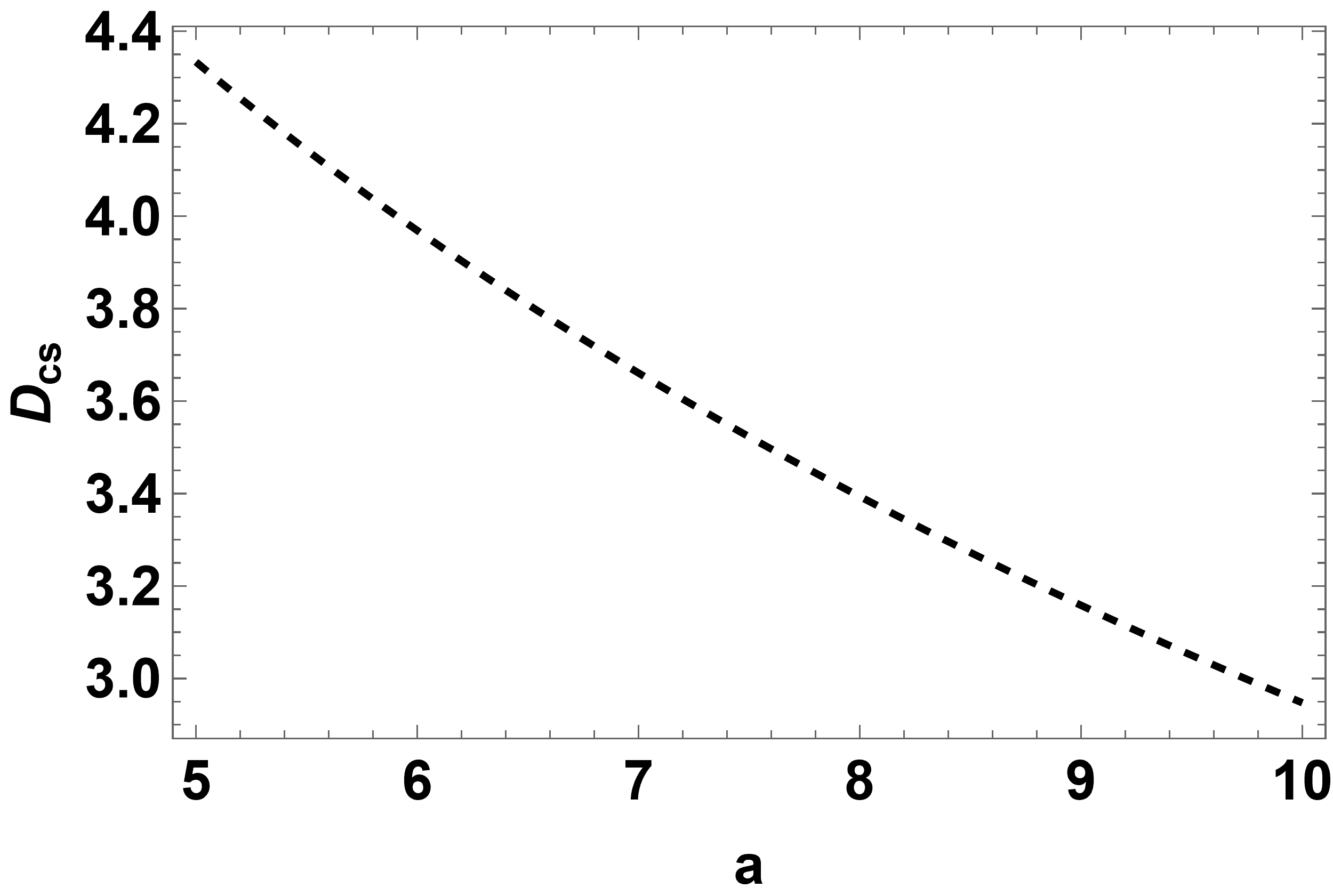}
  \caption{}
  \label{fig:7d}
\end{subfigure}
\begin{subfigure}{0.40\textwidth}
  \centering
  \includegraphics[width=\linewidth]{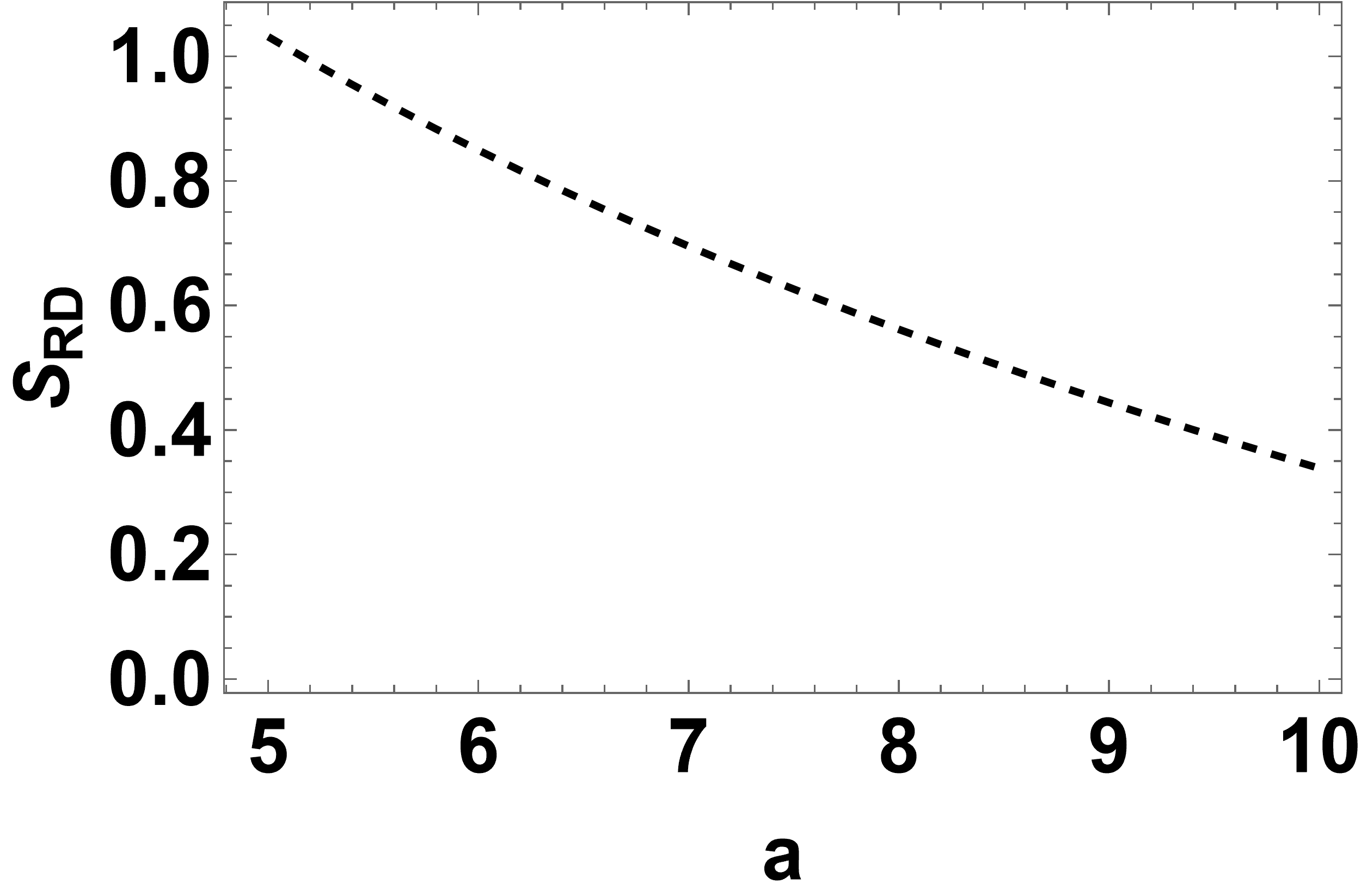}
  \caption{}
  \label{fig:7e}
\end{subfigure}\hfill
\begin{subfigure}{0.40\textwidth}
  \centering
  \includegraphics[width=\linewidth]{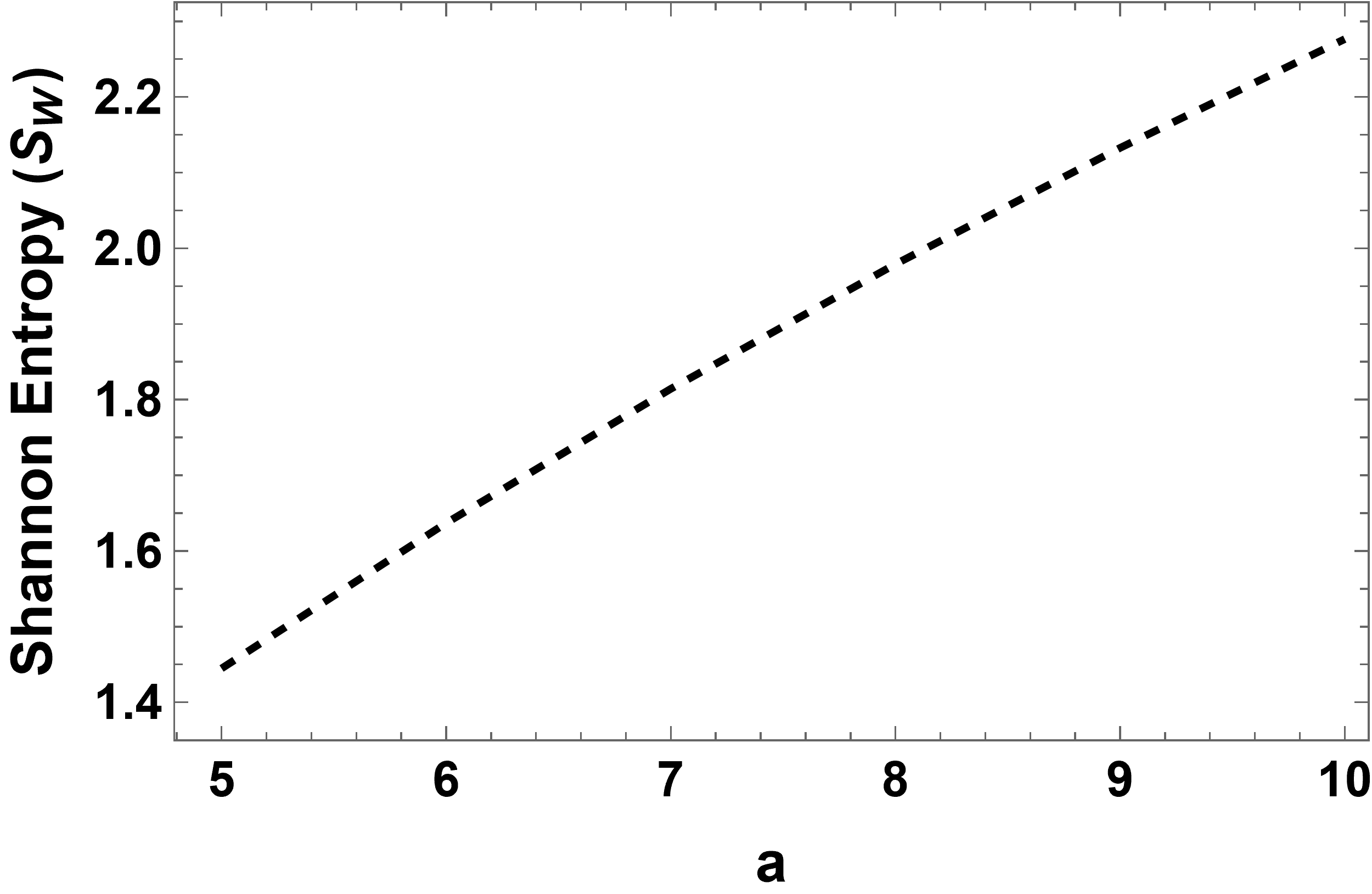}
  \caption{}
  \label{fig:7g}
\end{subfigure}
\caption{Plot~(\ref{fig:7a}) shows the Wehrl entropy $S_{H}$ of the Husimi distribution and Plot~(\ref{fig:7g}) shows the Shannon entropy obtained from the Wigner distribution. Plots~(\ref{fig:7b}), (\ref{fig:7c}) and (\ref{fig:7d}) show, respectively, the Kullback--Leibler divergence $S_{KL}$, the survival Jeffreys-type measure $\mathcal{R}$ [Eq.~(\ref{eq:Rsurvival})], and the Cauchy--Schwarz divergence
$D_{CS}$, all evaluated on the position-space marginals of the Wigner and Husimi distributions. Plot~(\ref{fig:7e}) shows the R\'enyi divergence $S_{RD}^{(\alpha)}$ [Eq.~(\ref{eq:Renyi_div})]
at $\alpha = 4$, computed from the position space and momentum space marginals respectively. All quantities are plotted as a function of the scattering length $a \times 10^{-10}\,\mathrm{m}$ at $N=10$.}
\label{fig:divergences}
\end{figure}

We have derived the phase space distributions and their corresponding marginals. The next natural question is how similar these distributions and their marginals actually are. Specifically, we will examine the degree of similarity or dissimilarity between the Wigner and Husimi distributions and explore how this resemblance changes with the scattering length. As the scattering length increases, interaction-induced broadening modifies the phase space structure, leading to larger quantitative deviations between the Wigner and Husimi representations. Since the Husimi distribution smooths fine scale features through Gaussian convolution, the divergence measures increase when interaction-driven redistribution of phase space density becomes more pronounced. In plot ~\ref{fig:7e}, it is clearly seen that as the value of $a$ increases, the resemblance between marginals obtained from both the distributions increases.  As the scattering length $a$ increases, the interactions between particles become stronger. Because the Husimi distribution smooths over these fine details, the difference between the two distributions increases. The trend of the Jeffreys divergence (see Fig.~\ref{fig:7b}) closely follows that of the relative (Kullback–Leibler) divergence. This similarity arises because both measures depend on the cumulative differences in entropy between the underlying probability distributions. In other words, since the Jeffreys divergence is a symmetrized form of the KL divergence, it captures comparable variations in the distributions, informational content, leading to analogous overall behavior. In Fig.~\ref{fig:7d}, we present the CS divergences, whose trend is opposite to that of the KL and Jeffreys divergences. The magnitude of CS divergence decreases as the value of scattering length increases. In Fig.~\ref{fig:7e}, the Rényi divergence is shown. The Rényi divergence quantifies the difference between the Rényi entropies derived from the Wigner and Husimi distributions. In general, the Rényi divergence increases when the scattering length $(a)$ increases. Thus, it exhibits a trend similar to that of the KL and Jeffreys divergences. The divergence measures shown in Fig.~\ref{fig:divergences} are evaluated using the position space marginals of the Wigner and Husimi distributions. The opposite trends observed between the KL/Jeffreys divergences and the Cauchy–Schwarz divergence arise from their different mathematical sensitivities. The KL and Jeffreys measures are logarithmic relative entropy distances and are particularly sensitive to local deviations in the tails of the distributions. In contrast, the Cauchy–Schwarz divergence depends on the global overlap between the two densities. As the interaction strength increases, the marginals broaden and their tails become more pronounced. This can enhance logarithmic deviations (increasing KL/Jeffreys) while simultaneously increasing global overlap (reducing CS divergence), leading to the opposite monotonic behavior. 
\subsection{Correlation measures}
\begin{figure}[H]
\centering
\begin{subfigure}{0.45\textwidth}
  \centering
  \includegraphics[width=\linewidth]{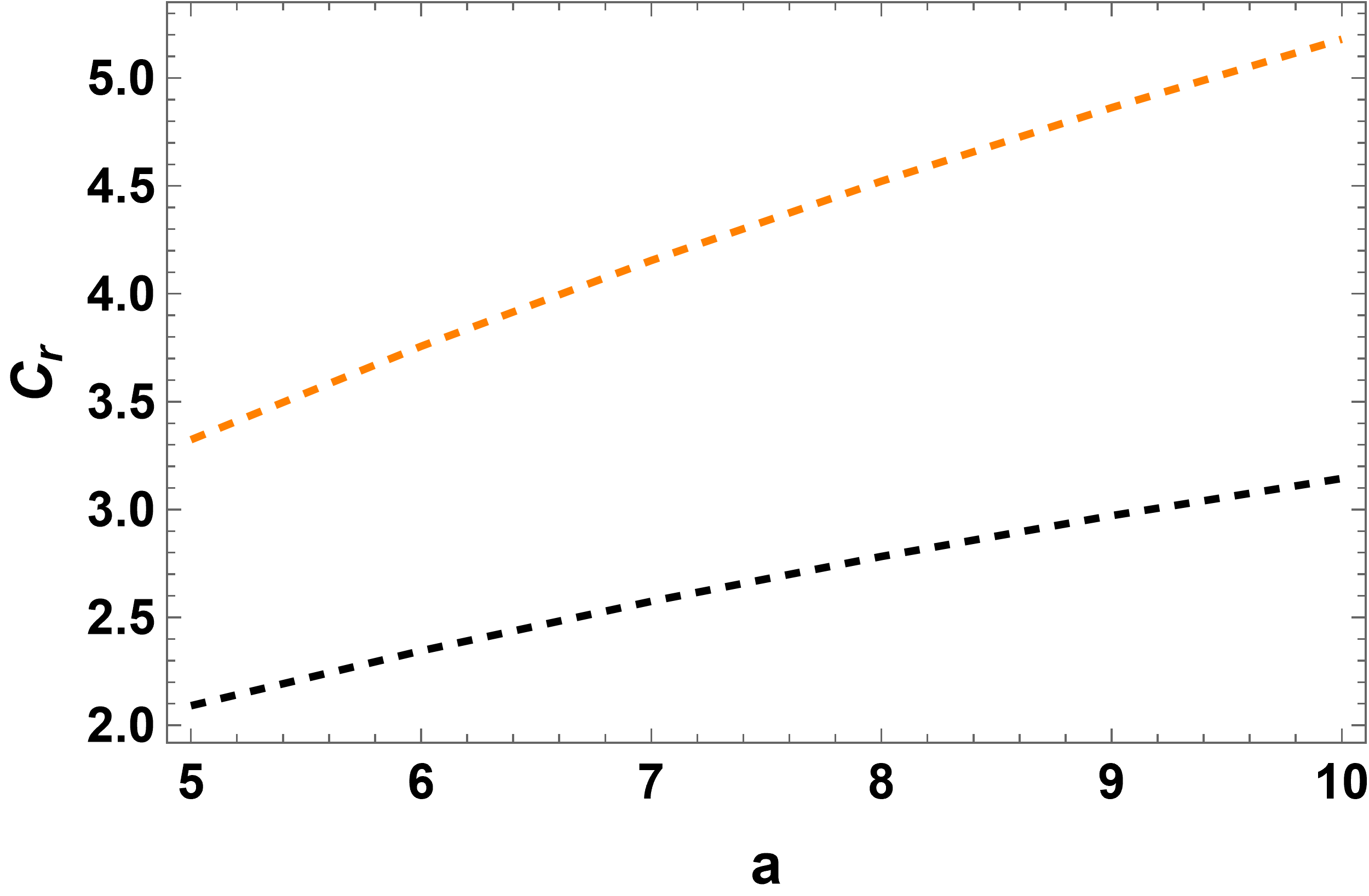}
  \caption{}
  \label{fig:ii55}
\end{subfigure}\hfill
\begin{subfigure}{0.45\textwidth}
  \centering
  \includegraphics[width=\linewidth]{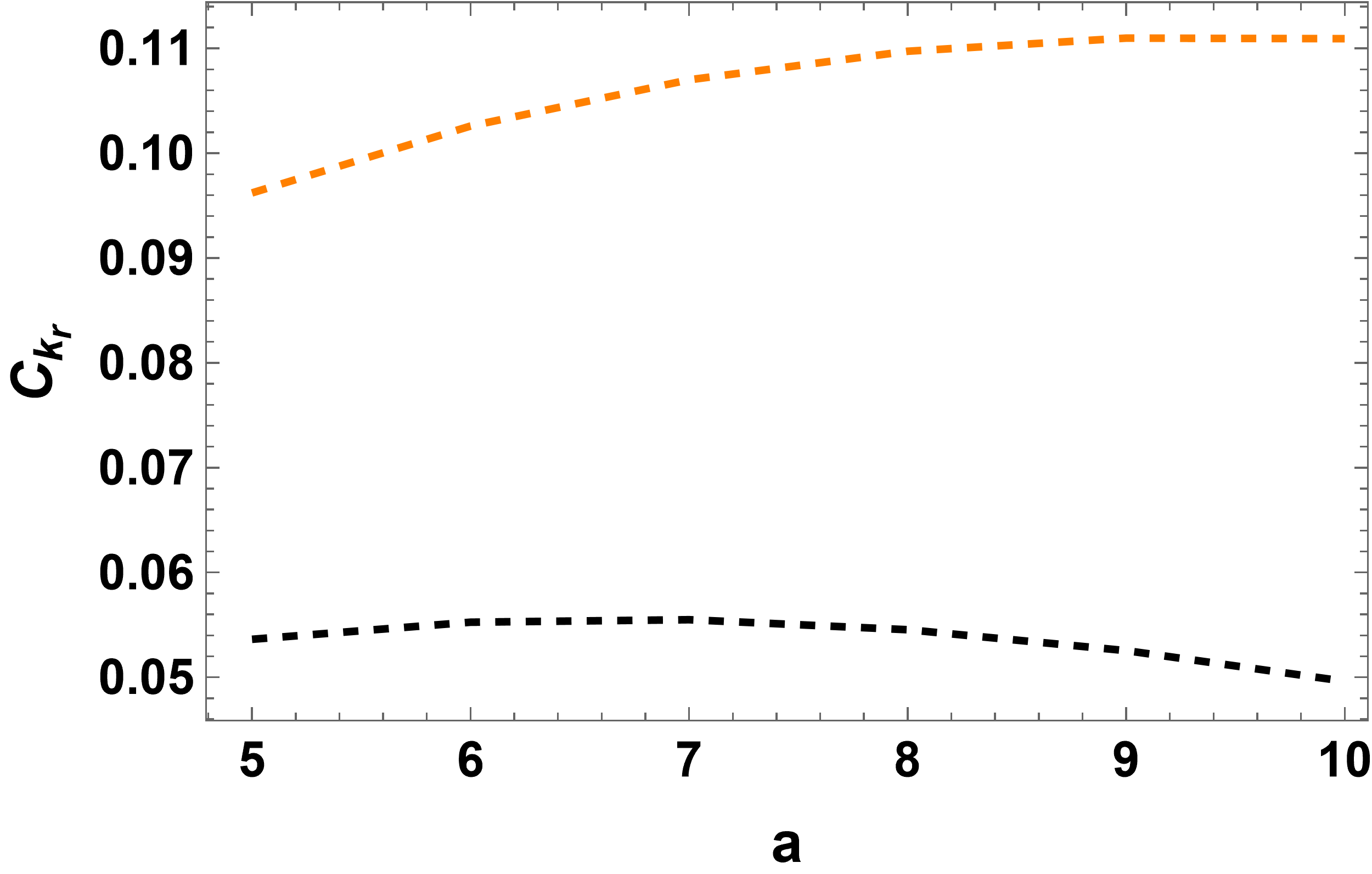}
  \caption{}
  \label{fig:ii11055}
\end{subfigure}\hfill
\begin{subfigure}{0.45\textwidth}
  \centering
  \includegraphics[width=\linewidth]{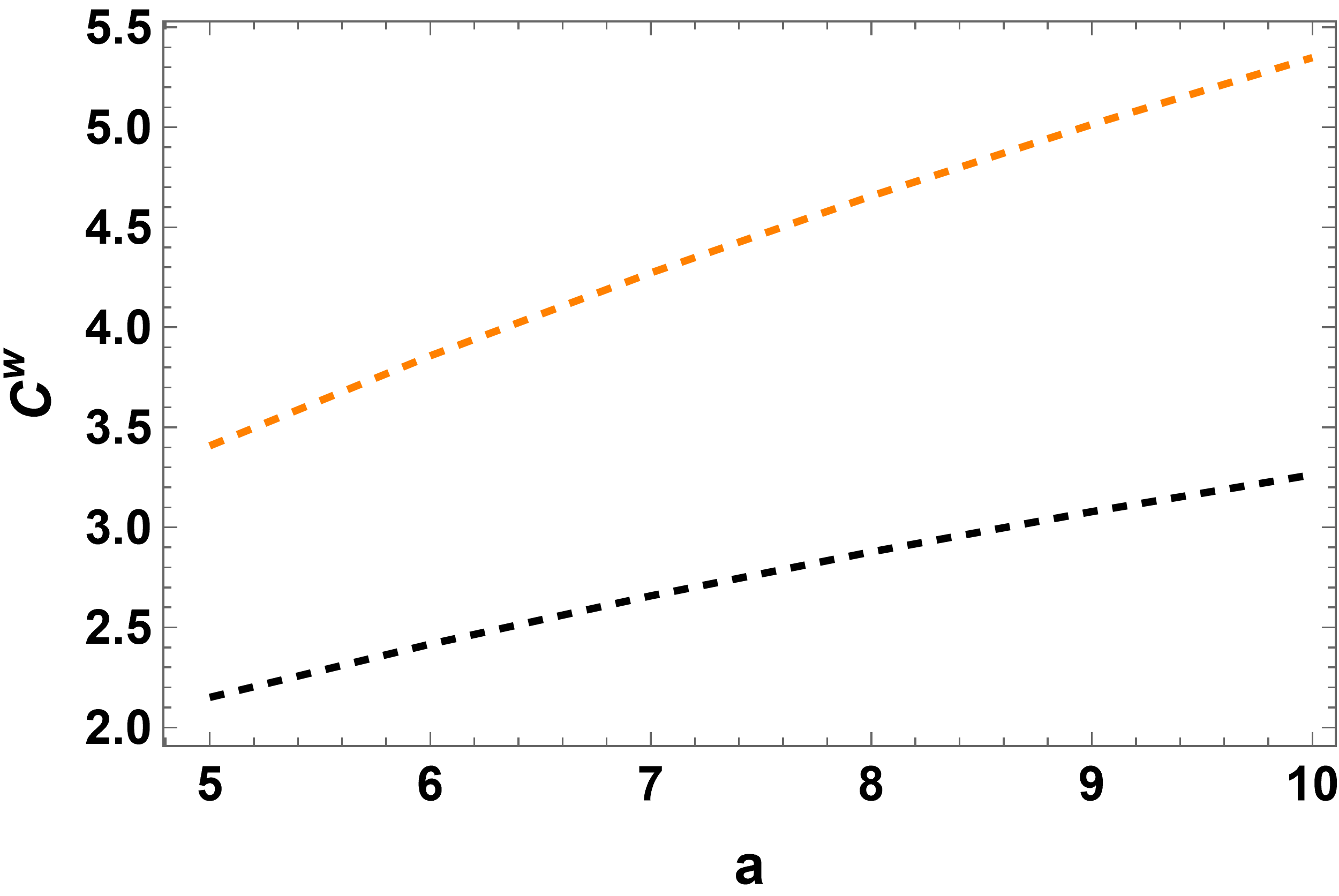}
  \caption{}
  \label{fig:98u}
\end{subfigure}\hfill
\begin{subfigure}{0.45\textwidth}
  \centering
  \includegraphics[width=\linewidth]{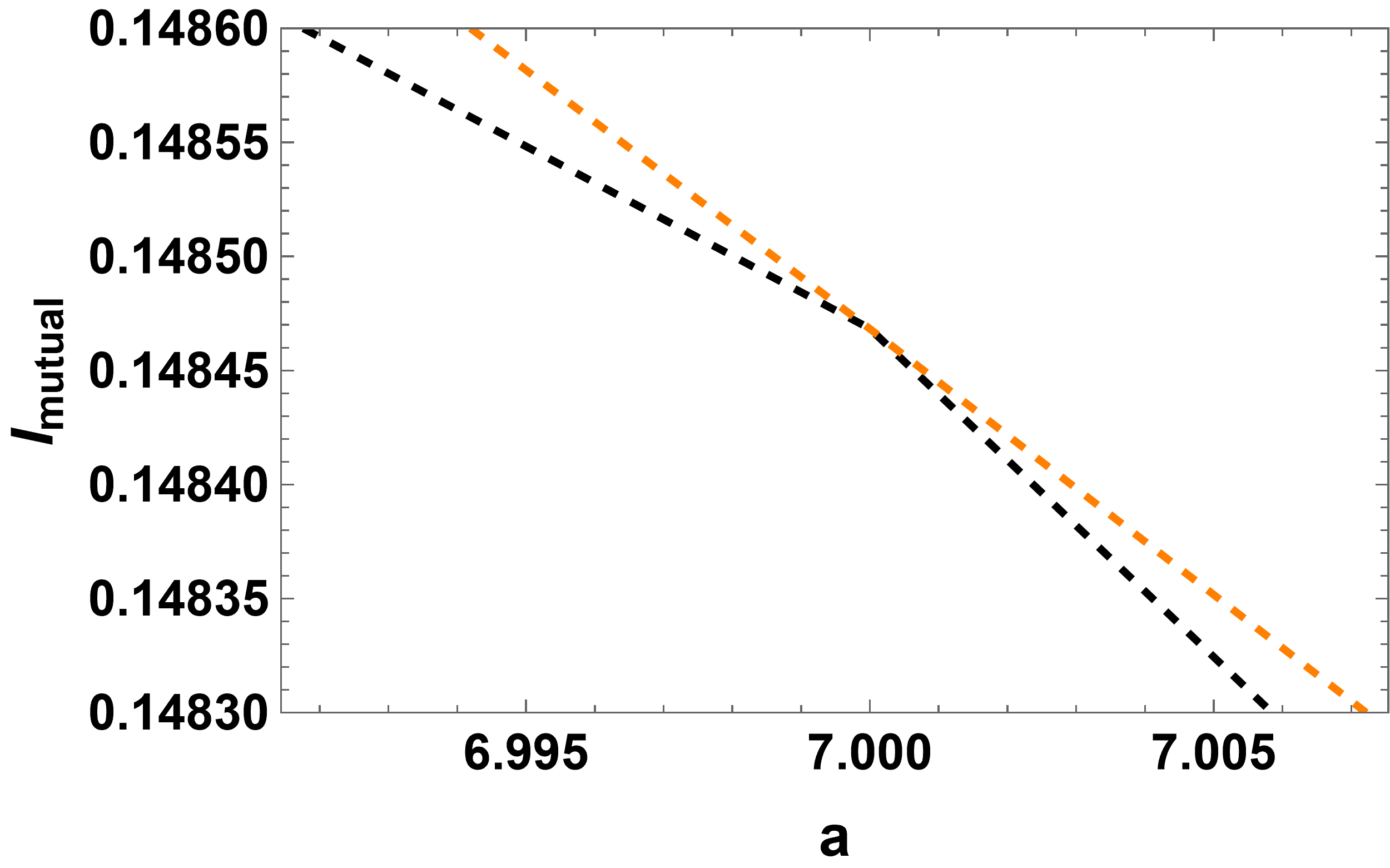}
  \caption{}
  \label{fig:54u}
\end{subfigure}
\caption{Plots~(~\ref{fig:ii55}) and (~\ref{fig:ii11055}) show the cumulative residual entropy $\mathcal{C}_{r}$ and $\mathcal{C}_{k_{r}}$ in position and momentum space respectively. Plot~(\ref{fig:98u}) shows the cross-cumulative residual entropy $\mathcal{C}^{W}_{r}$. They are plotted over the parameter range $a \in [5,10] \times 10^{-10}\,\mathrm{m}$
at $N = 10$. Plot~(\ref{fig:54u}) shows the mutual information $I_{\mathrm{mutual}}$ and is plotted over a selected parameter range of $a\in [5,10] \times 10^{-10}\,\mathrm{m}$ at $N = 10$. Note that the absolute variation of $I_{\mathrm{mutual}}$ over this range is small compared to the variations seen in the other information measures; the meaningful features are the monotonic decrease with $a$ and the systematic offset between the Wigner (black)
and Husimi (orange) curves, which converge at larger interaction strength. Results derived from the Wigner and Husimi distributions are indicated by black and orange lines, respectively.}
\label{fig:crosssumulative}
\end{figure}

In the previous sections, we discussed the phase space distributions, their marginals, their entropies and the corresponding divergences. These aspects are quantified by examining correlation measures, which depend on the properties of the distributions and their associated entropies. We begin by discussing the cumulative residual entropy, which is the entropy corresponding to survivals. This cumulative residual entropy shows distinct features. The magnitude of cumulative residual entropy increases as the scattering length increases in both position and momentum space. This entropy, shown in plots ~\ref{fig:ii55} and ~\ref{fig:ii11055}, provides an alternative perspective on the uncertainty or spread in a system, emphasizing the contribution of the survival probabilities rather than that of the entire distribution. We presented cross cumulative residual entropy in Fig. ~\ref{fig:98u}. The magnitude of it increases with the increase in scattering length $(a)$. The numerical value of the entropy obtained from the Wigner distribution is lesser because it includes quantum interference effects, which are absent in the smoothed Husimi distribution. In the parameter regime considered here, since the Wigner distribution remains positive definite, the mutual information is strictly real valued, and no imaginary component arises in the numerical evaluation. The opposite trends of $\mathcal{C}^{W}_{r}$ and $I_{\mathrm{mutual}}$
deserve a brief comment, since they are not contradictory but reflect the fact that the two quantities measure different aspects of the joint phase space distribution. The cross cumulative residual entropy $\mathcal{C}^{W}_{r}$ quantifies the total residual phase space uncertainty after subtracting the purely positional contribution; it increases with $a$ because interaction-induced broadening enlarges the support of the joint distribution in phase space, leaving more ``room'' for residual uncertainty. The mutual information
$I_{\mathrm{mutual}}$, by contrast, measures how much information is shared between $r$ and $k_{r}$, normalized by the product of the marginals. As interactions broaden the joint distribution, $W$ approaches $\rho(r)\,\rho(k_{r})$ more closely --- the joint
distribution factorizes more, and the shared information decreases even as the total uncertainty increases. In short, $\mathcal{C}^{W}_{r}$ tracks total residual uncertainty while $I_{\mathrm{mutual}}$ tracks position--momentum correlation;
broadening simultaneously increases the support and reduces the coupling between conjugate variables, and the two measures therefore move in opposite directions.

In Fig.~\ref{fig:54u} we show that the mutual information decreases as the scattering length increases. The gradual decrease of mutual information with increasing scattering length indicates that stronger interactions lead to a broader phase space distribution and reduced position–momentum correlations. The Husimi-based mutual information remains systematically larger than the Wigner case due to the Gaussian coarse-graining inherent in the Husimi representation, which enhances classical correlations by suppressing quantum interference structures. The convergence of the two curves at larger scattering length suggests a transition toward a more semiclassical phase space structure, where the difference between the Wigner and Husimi descriptions becomes less pronounced. In the present work, the mutual information is defined between the conjugate phase space variables $r$ and $k$ rather than between spatially separated particles or distinct many-body subsystems. Specifically, it is computed from the joint phase space distribution and its corresponding marginals, thereby quantifying the statistical dependence between position and momentum degrees of freedom within the reduced one-body description. Since the Gross–Pitaevskii framework incorporates interaction effects through a nonlinear mean field wavefunction, the resulting mutual information captures interaction-induced phase space correlations and redistribution of coherence, rather than genuine particle–particle entanglement. Consequently, throughout this work, mutual information should be interpreted as a measure of phase space correlation structure and information sharing between conjugate variables, providing insight into how interactions reshape localization and delocalization in the system. It is important to distinguish this quantity from the mutual information used in quantum information science, where the relevant quantity is typically the von Neumann mutual information between spatially separated subsystems or between distinguishable particles, and serves as a measure of quantum entanglement or classical correlations between those subsystems. The mutual information computed here is qualitatively different in two respects. First, it is defined between the position and momentum degrees of freedom of a single effective particle — the conjugate variables of the reduced one-body phase space — rather than between two spatially separated or physically distinct subsystems. Second, because the Gross-Pitaevskii framework is a mean-field theory that describes the condensate as a pure product state at the many-body level, it does not in principle support genuine many-body entanglement. Any entanglement between particles is averaged out in the mean-field approximation, and the resulting single-body Wigner or Husimi function carries no information about inter-particle correlations beyond what is encoded in the effective nonlinear potential. Consequently, the mutual information and related measures computed in this work should not be interpreted as entanglement measures in the quantum information theoretic sense, and we do not claim that they are directly applicable to quantum communication or computation protocols, which rely on particle-particle entanglement rather than phase space correlations. What these measures do capture is the interaction-induced restructuring of the phase space distribution — specifically, the degree to which position and momentum are statistically dependent in the evolved condensate state, and this restructuring has well-defined physical consequences for delocalization, coherence loss, and the semiclassical transition as interaction strength increases. The connection to quantum thermodynamics is more direct: entropic measures of phase space spreading are naturally related to thermodynamic work, irreversibility, and the informational entropy generated during the transition from a coherent quantum state to a more classical mixed description, and it is in this context that the present framework has the most immediate relevance.

\section{Concluding remarks}\label{sec:conclusion}
In the present study, using ultracold atomic gases as an illustration, we demonstrate that the quantum characteristics of the system decrease when transitioning from the Wigner to the Husimi representation. In summary, the collective behavior of the various information-theoretic measures reveals how interaction strength governs the redistribution of uncertainty and information in many-body quantum systems. The analysis of the Wigner and Husimi distributions for different scattering lengths $(a)$ and particle numbers $(N)$ reveals how quantum coherence manifests in phase space and how it is gradually suppressed under coarse graining. In general, negativity of the Wigner distribution is associated with quantum interference and nonclassical correlations. However, in the present ground state configuration, the Wigner function remains positive definite, and the quantum features discussed here arise from interaction-driven phase space restructuring rather than explicit negativity. The observed broadening of the phase space distributions with increasing interaction strength originates from repulsive interparticle interactions in the Gross–Pitaevskii framework, which introduce an interaction-dependent length scale that spreads the condensate in real space and suppresses fine-scale coherence in phase space. Differences between the Wigner and Husimi-based survival functions reflect the influence of phase space smoothing in the Husimi representation. While marginal survival functions do not directly display Wigner negativity, the quantitative deviations between the two distributions encode how coarse-graining modifies the underlying phase space structure. The increase of Shannon and Wehrl entropies with the scattering length $a$ indicates enhanced phase space delocalization due to stronger repulsive interactions and increasing many-body correlations ~\cite{wehrl1978general}.  The comparison of full and marginal survival functions further demonstrates that true delocalization requires simultaneous spreading in both position and momentum. Overall, the results illustrate the transition from quantum to classical behavior as one moves from the Wigner to the Husimi representation and as the scattering length increases.

Concurrently, the Fisher information shows the opposite trend, increasing in position space and decreasing in momentum space, consistent with the global uncertainty. The Rényi entropy, sensitive to non-Gaussian features, decreases with increasing $a$, signifying the decrease of accessible phase space volume and reduction of effective complexity ~\cite{bialynicki2006formulation}. Divergence measures such as the Kullback–Leibler, Jeffreys, and Cauchy–Schwarz divergences further quantify how the Husimi distribution deviates from the Wigner function as interaction increases, which results in the loss of quantum interference in the phase space description. In short, the monotonic increase of KL and Jeffreys divergences with interaction strength indicates that stronger many-body correlations generate increasingly non-Gaussian phase space structures that cannot be captured by coarse-grained Husimi representations. Physically, this reflects interaction-induced coherence loss and fragmentation of phase space structure, rather than simple statistical broadening. From an information-theoretic perspective, the cumulative residual and cross-cumulative residual entropies together capture how uncertainty and correlation evolve in interacting many-body systems. The cumulative residual entropy, which emphasizes survival probabilities, increases with increasing scattering length, indicating that stronger interparticle interactions increase random fluctuations and decrease spatial ordering, thereby enhancing the residual uncertainty of the system. Similarly, the cross-cumulative residual entropy increases with the increase in scattering length, reflecting the growth of interaction-driven redistribution of phase space density, reflecting the nonlinear mean-field coupling rather than genuine inter-particle correlations beyond the GP approximation, as the system becomes more strongly interacting. The mutual information decreases slightly with increasing scattering length, indicating that stronger interactions broaden the phase space distribution and reduce position-momentum correlations. The convergence of the Wigner and Husimi mutual information at larger scattering length indicates that the phase space structure becomes increasingly semiclassical, with reduced quantum interference effects.

In conclusion, the present results are not only of theoretical significance but also hold potential relevance for a wide range of practical applications. The analysis of phase space distributions, in particular, is expected to contribute to the fields of quantum communication, quantum computation, and, with particular relevance to quantum thermodynamics, where phase space entropic measures characterize coherence loss, irreversibility, and the energetic cost of quantum-to-classical transitions in interacting many-body systems~\cite{deffner2025quantum}. In quantum communication, these measures provide a means to quantify information capacity and decoherence in noisy or strongly correlated environments. In quantum computation, they are helpful to characterize entanglement structure, information retention, and state delocalization in multi-qubit and continuous variable systems. In quantum thermodynamics, the entropic and divergence-based quantities established here naturally relate to the generation of informational entropy, energetic cost, and the transition from quantum coherence to classical behavior in interacting systems. 

A natural direction in which the present framework connects to broader themes in quantum information is the resource-theoretic role of Wigner negativity, also known as ``MAGIC'' in the context of quantum
computation. In continuous variable quantum optics, negativity of the Wigner function has been established as a necessary resource for any super-polynomial quantum computational speed up \cite{MariEisert2012,VeitchFerrieEmerson2012}, and the recent proof that contextuality and Wigner negativity are equivalent notions of non-classicality for standard models of continuous-variable quantum computing \cite{BoothEmeriauKashefiMansfield2022} has placed
phase space negativity at the centre of the resource-theoretic description of quantum advantage. A formal resource theory of Wigner
negativity, together with a computable monotone (the Wigner logarithmic negativity), has been developed by Albarelli, Genoni, Paris and Ferraro \cite{AlbarelliGenoniParisFerraro2018} and analogous
notions of magic have been formulated for both qudit and bosonic encodings, including the Gottesman--Kitaev--Preskill code \cite{HowardWallmanVeitchEmerson2014, GarciaAlvarez2020}. In the present work, the Gross Pitaevskii ground state of a harmonically
trapped condensate has a positive-definite Wigner function over the parameter range explored, so the system carries no Wigner negativity and, in this sense, no ``MAGIC'': the interaction-induced
restructuring we identify is encoded in the redistribution of a positive phase-space density rather than in the appearance of negative
regions. Within this regime, the entropic and divergence-based measures we have computed quantify how interactions reshape coherence and complexity at zero negativity. It is precisely for this reason that we expect the present framework to provide a useful baseline for studies that move beyond the positive-definite regime, where Wigner negativity does emerge --- for instance, in attractive
condensates near the collapse instability, in strongly correlated states beyond the Gross--Pitaevskii approximation, or in driven systems where non-Gaussian dynamics can generate Wigner-negative features \cite{ChabaudEmeriauGrosshans2021}. In such regimes, the
Shannon and Wehrl entropies, the cumulative residual entropies and the divergence measures introduced here would naturally complement direct quantifiers of negativity (such as the Wigner logarithmic
negativity) by characterizing the entropic and statistical structure of the underlying phase-space distribution, and the Wigner and Husimi comparison would acquire genuinely independent content beyond the
fixed Gaussian-smoothing offset observed here. Establishing how the information theoretic measures developed in this work behave in the presence of Wigner negativity and, in particular, whether they can
serve as proxies or witnesses for magic in interacting many body systems is an interesting direction for future work. Overall, this work presents an information-theoretic framework to describe the evolution of correlations, coherence, and uncertainty in complex quantum systems. Further extensions of this approach to open, non-equilibrium, and higher-dimensional systems may further strengthen the connection between quantum information theory and practical quantum technologies.

\subsubsection*{Acknowledgements}
The authors thank the anonymous referee for valuable comments and suggestions that helped improve the manuscript. 

\subsubsection*{Competing Interests}
The authors have equal contributions and declare no competing financial or personal interests.

\appendix
\section{Wave function calculation}\label{sec:appendix}
The non-linear Schrodinger equation for BEC gases is given by Gross Pitaevski equation
\begin{equation}
    i\hbar \frac{\partial \psi}{\partial t} = \Bigg[-\frac{\hbar^{2}}{2m}\nabla^{2}+g|\psi(r)|^2+V(r)\Bigg]\psi(r).
\end{equation}
We work in natural units, $\hbar = c = 1$, and our potential is a three-dimensional harmonic trap.  Thus, the equation modifies as follows
\begin{equation}
    i \frac{\partial \psi(r,t)}{\partial t} = \Bigg[-\frac{1}{2m}\nabla^{2}+\kappa_{t}|\psi(r,t)|^2+V(r)\Bigg]\psi(r,t),
\end{equation}
such that $\int d^{3}r\, |\psi|^{2} = 1$, $\kappa_{t}$ is the nonlinear term, and 
$$V(r) = \frac{1}{2}m\omega^{2}(\lambda^{2}_{x}x^{2}+\lambda^{2}_{y}y^{2}+\lambda^{2}_{z}z^{2}).$$
Assuming $\lambda_{x} = \lambda_{y} = 1$, and choosing cylindrical coordinates, the potential changes as
$$V(r) = \frac{1}{2}m\omega^{2}(\rho^{2}+\lambda^{2}_{z}z^{2}).$$
With the following change of variables,
$$\tau = \omega_{r}t,~ a_{0}\rho = r,~ a_{0}s = z,~ Q = -\frac{8\pi\,a\, N}{a_{0}},~ a_{0} = \frac{1}{r}\,\sqrt{\frac{1}{m\omega_{r}}},~ \lambda_{z} = \frac{\omega_{z}}{\omega_{r}},$$
The wave function and Gross-Pitaevskii equation are modified to
$$u(\rho,\, s,\, \tau) = \psi(r,\, z,\, ,t)\sqrt{\frac{a_{0}^{3}}{N}},$$ and
\begin{equation}
    i\, \frac{\partial u}{\partial \tau} = \Bigg[-\frac{1}{2}\nabla^{2} + \frac{1}{2}\bigg(\rho^{2} + \lambda^{2}_{z}s^{2}\bigg) - \frac{Q}{2}|u|^{2}\Bigg]u(\rho,\,s,\, \tau),
\end{equation}
such that $\int |u|^{2} dr = 1$. Using the separation of variables
$u(\rho,\, s,\, \tau) = \phi(\rho)\xi(s,\tau)$ gives
\begin{equation}
    -\frac{1}{2}\nabla^{2}\phi + \frac{1}{2}\rho^{2}\phi = \omega_{\rho}\phi.
\end{equation}
This is a well-known eigenvalue problem whose ground state solution is given by
\begin{equation}
    \phi_{0}(\rho) = \exp{\bigg[-\frac{\rho^{2}}{2}\bigg]}
\end{equation}
Finally, we do a transformation, 
\begin{equation}
    \varphi(s,\, \tau) = \xi(s)e^{i\omega_{\rho}\tau},
\end{equation}
resulting in
\begin{equation}
    2i\, \frac{\partial \varphi}{\partial \tau} +\frac{\partial^{2}\varphi}{\partial s^{2}}+\lambda_{z}^{2}s^{2}\varphi -\frac{Q}{2}|\varphi|^{2}\varphi = 0.
\end{equation}
Now, we put $\lambda_{z}= 0$ and so 
\begin{equation}
    2i\, \frac{\partial \varphi}{\partial \tau} +\frac{\partial^{2}\varphi}{\partial s^{2}} -\frac{Q}{2}|\varphi|^{2}\varphi = 0.
\end{equation}
This is a one-dimensional NLSE, and it has the following solution for the stationary part 
\begin{equation}
    \varphi(s) = \frac{\sqrt{Q}}{4\pi} sech\bigg(\frac{Q\,s}{8\pi}\bigg).
\end{equation}
This implies
 \begin{equation}
     u(\rho,\, s,\, \tau) = \frac{\sqrt{Q}}{4\pi} sech\bigg(\frac{Q\,s}{8\pi}\bigg) e^{-\frac{\rho^{2}}{2}}e^{i\, \omega_{\rho}\tau}.
 \end{equation}
 Substituting back the initial variables
 $\rho = \frac{r}{a_{0}}, s = \frac{z}{a_{0}}, \tau = \omega_{r}t$, we get
 \begin{equation}
     \psi(r,z,t) = \mathcal{N}\,\frac{i}{4\pi}\sqrt{\frac{8\pi\,a\,N}{a_{0}}}sech\bigg(\frac{aN}{a_{0}^{2}}z\bigg)\exp{\bigg[-\frac{r^{2}}{2a_{0}^{2}}\bigg]}e^{i\omega_{r}t}.
 \end{equation}
 where $\mathcal{N}$ is the normalization constant. Upon normalizing, we obtained the normalized wave function as
 \begin{align}
     \psi(r,z,t) & = \frac{i}{4\pi\, a_{0}^{2}}\sqrt{8\pi\,a\,N}\,sech\bigg(\frac{aN}{a_{0}^{2}}z\bigg)\exp{\bigg[-\frac{r^{2}}{2a_{0}^{2}}\bigg]}e^{i\omega_{r}t}.
 \end{align}
The Fourier transform of this wave function gives us the wave function in momentum space. After normalization, we get
 \begin{align}
     \phi(k_{r},k_{z},t) & = \frac{i\,\pi\, a_{0}^{2}}{\sqrt{2\pi\,a N}}sech\bigg(\frac{\pi\,a_{0}^{2}}{2aN}k_{z}\bigg)\exp{\bigg[-\frac{1}{2}k^{2}_{r}a^{2}_{0}\bigg]}e^{i\omega_{r}t}
 \end{align}
where $a_{0}$ is the oscillator length, $N$ is the mean number of bosons that are trapped, $a$ scattering length and $\kappa = \frac{4\pi\, a N}{a_{0}}$. In this entire system, the axial direction is fixed. In other words, the value of $z$ is fixed. So, the normalized wave function in position and momentum space is given by 
\begin{align}
    \psi(r,t) & = \frac{i}{4\pi a^{2}_{0}}\sqrt{8\pi\,a\,N}sech\bigg(\frac{aN}{a_{0}^{2}}z\bigg)\exp{\bigg[-\frac{r^{2}}{2a_{0}^{2}}\bigg]}e^{i\omega_{r}t},
\end{align}
\begin{align}
     \phi(k_{r},t) & = \frac{i\,\pi\, a_{0}^{2}}{\sqrt{2\pi\,a N}}sech\bigg(\frac{\pi\,a_{0}^{2}}{2aN}k_{z}\bigg)\exp{\bigg[-\frac{1}{2}k^{2}_{r}a^{2}_{0}\bigg]}e^{i\omega_{r}t}.
 \end{align}
Using this, we get Wigner and Husimi distributions as follows
\begin{align}
W(r, k_{r}) & = 
\frac{\mathcal{A}(z)}{\pi}\,
\exp\!\left(-\frac{r^{2}}{a_{0}^{2}}\right)
\exp\!\left(-k_{r}^{2}\,a_{0}^{2}\right),
\end{align}
and the Husimi distribution, obtained by convolution with a Gaussian kernel of unit variance in both arguments, is
\begin{align}
H(r, k_{r}) & =
\frac{\mathcal{A}(z)}{\pi\sqrt{(a_{0}^{2}+1)(a_{0}^{-2}+1)}}\,
\exp\!\left(-\frac{r^{2}}{a_{0}^{2}+1}\right)
\exp\!\left(-\frac{k_{r}^{2}}{a_{0}^{-2}+1}\right),
\end{align}
where 
$$\mathcal{A}(z) =
\frac{8\pi a N}{(4\pi a_{0}^{2})^{2}}\,
\mathrm{sech}^{2}\!\left(\frac{a N}{a_{0}^{2}}\,z\right).$$
\bibliographystyle{unsrt}
\bibliography{ref1}

\begin{thebibliography}{10}

\bibitem{jeon2015introduction}
Sangyong Jeon and Ulrich Heinz.
\newblock Introduction to hydrodynamics.
\newblock {\em International Journal of Modern Physics E}, 24(10):1530010, 2015.

\bibitem{mendoncca2008collective}
JT~Mendon{\c{c}}a, R~Kaiser, H~Ter{\c{c}}as, and J~Loureiro.
\newblock Collective oscillations in ultracold atomic gas.
\newblock {\em Physical Review A—Atomic, Molecular, and Optical Physics}, 78(1):013408, 2008.

\bibitem{phenix2010azimuthal}
PHENIX Collaboration et~al.
\newblock Azimuthal anisotropy of neutral pion production in au+ au collisions at sqrt (s\_nn)= 200 gev: Path-length dependence of jet quenching and the role of initial geometry.
\newblock {\em arXiv preprint arXiv:1006.3740}, 2010.

\bibitem{martinez2012boost}
Mauricio Martinez, Radoslaw Ryblewski, and Michael Strickland.
\newblock Boost-invariant (2+ 1)-dimensional anisotropic hydrodynamics.
\newblock {\em Physical Review C—Nuclear Physics}, 85(6):064913, 2012.

\bibitem{voloshin2002anisotropic}
SA~Voloshin.
\newblock Anisotropic flow.
\newblock {\em arXiv preprint nucl-ex/0210014}, 2002.

\bibitem{mukherjee2015wigner}
Asmita Mukherjee, Sreeraj Nair, and Vikash~Kumar Ojha.
\newblock Wigner distributions for gluons in a light-front dressed quark model.
\newblock {\em Physical Review D}, 91(5):054018, 2015.

\bibitem{mukherjee2014quark}
Asmita Mukherjee, Sreeraj Nair, and Vikash~Kumar Ojha.
\newblock Quark wigner distributions and orbital angular momentum in light-front dressed quark model.
\newblock {\em Physical Review D}, 90(1):014024, 2014.

\bibitem{giorgini2008theory}
Stefano Giorgini, Lev~P Pitaevskii, and Sandro Stringari.
\newblock Theory of ultracold atomic fermi gases.
\newblock {\em Reviews of Modern Physics}, 80(4):1215--1274, 2008.

\bibitem{bloch2008many}
Immanuel Bloch, Jean Dalibard, and Wilhelm Zwerger.
\newblock Many-body physics with ultracold gases.
\newblock {\em Reviews of modern physics}, 80(3):885--964, 2008.

\bibitem{leggett2001bose}
Anthony~J Leggett.
\newblock Bose-einstein condensation in the alkali gases: Some fundamental concepts.
\newblock {\em Reviews of modern physics}, 73(2):307, 2001.

\bibitem{dalfovo1999theory}
Franco Dalfovo, Stefano Giorgini, Lev~P Pitaevskii, and Sandro Stringari.
\newblock Theory of bose-einstein condensation in trapped gases.
\newblock {\em Reviews of modern physics}, 71(3):463, 1999.

\bibitem{gross2017quantum}
Christian Gross and Immanuel Bloch.
\newblock Quantum simulations with ultracold atoms in optical lattices.
\newblock {\em Science}, 357(6355):995--1001, 2017.

\bibitem{bloch2012quantum}
Immanuel Bloch, Jean Dalibard, and Sylvain Nascimbene.
\newblock Quantum simulations with ultracold quantum gases.
\newblock {\em Nature Physics}, 8(4):267--276, 2012.

\bibitem{kar2026thermodynamic}
Tiyasa Kar, Atul Kedia, and Ramkumar Radhakrishnan.
\newblock Thermodynamic characteristics of a fermi gas with an invariant energy scale and its astrophysical implications.
\newblock {\em arXiv preprint arXiv:2601.17004}, 2026.

\bibitem{ojha2025quantum}
Vikash~Kumar Ojha, Ramkumar Radhakrishnan, and Mariyah Ughradar.
\newblock Quantum information measures in quartic and symmetric potentials using perturbative approach.
\newblock {\em Physica A: Statistical Mechanics and its Applications}, 659:130346, 2025.

\bibitem{augusiak2012many}
R~Augusiak, FM~Cucchietti, and M~Lewenstein.
\newblock Many-body physics from a quantum information perspective.
\newblock In {\em Modern Theories of Many-Particle Systems in Condensed Matter Physics}, pages 245--294. Springer, 2012.

\bibitem{willby2025quantum}
Christopher Willby, Martin Kiffner, Joseph Tindall, Jason Crain, and Dieter Jaksch.
\newblock Quantum information perspective on many-body dispersive forces.
\newblock {\em Physical Review Letters}, 135(11):110403, 2025.

\bibitem{carter2007introduction}
Tom Carter.
\newblock An introduction to information theory and entropy.
\newblock {\em Complex systems summer school, Santa Fe}, pages 15--34, 2007.

\bibitem{gray2011entropy}
Robert~M Gray.
\newblock {\em Entropy and information theory}.
\newblock Springer Science \& Business Media, 2011.

\bibitem{bromiley2004shannon}
PA~Bromiley, NA~Thacker, and E~Bouhova-Thacker.
\newblock Shannon entropy, renyi entropy, and information.
\newblock {\em Statistics and Inf. Series (2004-004)}, 9(2004):2--8, 2004.

\bibitem{veyrat2009mutual}
Nicolas Veyrat-Charvillon and Fran{\c{c}}ois-Xavier Standaert.
\newblock Mutual information analysis: how, when and why?
\newblock In {\em International workshop on cryptographic hardware and embedded systems}, pages 429--443. Springer, 2009.

\bibitem{kraskov2004estimating}
Alexander Kraskov, Harald St{\"o}gbauer, and Peter Grassberger.
\newblock Estimating mutual information.
\newblock {\em Physical review E}, 69(6):066138, 2004.

\bibitem{behera2025mutual}
Satyajit Behera, Javier~E Contreras-Reyes, and Suchandan Kayal.
\newblock Mutual information matrix and global measure based on tsallis entropy.
\newblock {\em Nonlinear Dynamics}, 113(6):5239--5249, 2025.

\bibitem{csiszar2004information}
Imre Csisz{\'a}r, Paul~C Shields, et~al.
\newblock Information theory and statistics: A tutorial.
\newblock {\em Foundations and Trends{\textregistered} in Communications and Information Theory}, 1(4):417--528, 2004.

\bibitem{lin2002divergence}
Jianhua Lin.
\newblock Divergence measures based on the shannon entropy.
\newblock {\em IEEE Transactions on Information theory}, 37(1):145--151, 2002.

\bibitem{contreras2022renyi}
Javier~E Contreras-Reyes.
\newblock R{\'e}nyi entropy and divergence for varfima processes based on characteristic and impulse response functions.
\newblock {\em Chaos, Solitons \& Fractals}, 160:112268, 2022.

\bibitem{fisher1}
Ronald~A. Fisher.
\newblock On the mathematical foundations of theoretical statistics.
\newblock {\em Philosophical transactions of the Royal Society of London. Series A, containing papers of a mathematical or physical character}, 222(594-604):309--368, 1922.

\bibitem{falaye2016fisher}
Babatunde~J Falaye, Fernando~A Serrano, and Shi~H Dong.
\newblock Fisher information for the position-dependent mass schr{\"o}dinger system.
\newblock {\em Physics Letters A}, 380(1-2):267--271, 2016.

\bibitem{kirby2015feasibility}
BT~Kirby, GT~Hickman, TB~Pittman, and JD~Franson.
\newblock Feasibility of single-photon cross-phase modulation using metastable xenon in a high finesse cavity.
\newblock {\em Optics Communications}, 337:57--61, 2015.

\bibitem{rand2016lectures}
Stephen~Colby Rand.
\newblock {\em Lectures on light: nonlinear and quantum optics using the density matrix}.
\newblock Oxford University Press, 2016.

\bibitem{chang2014quantum}
Darrick~E Chang, Vladan Vuleti{\'c}, and Mikhail~D Lukin.
\newblock Quantum nonlinear optics—photon by photon.
\newblock {\em Nature Photonics}, 8(9):685--694, 2014.

\bibitem{rolston2002nonlinear}
SL~Rolston and William~D Phillips.
\newblock Nonlinear and quantum atom optics.
\newblock {\em Nature}, 416(6877):219--224, 2002.

\bibitem{wigner1932quantum}
Eugene Wigner.
\newblock On the quantum correction for thermodynamic equilibrium.
\newblock {\em Physical review}, 40(5):749, 1932.

\bibitem{radhakrishnan2022wigner}
Ramkumar Radhakrishnan and Vikash~Kumar Ojha.
\newblock Wigner distribution of sine-gordon and kink solitons.
\newblock {\em Modern Physics Letters A}, 37(37n38):2250236, 2022.

\bibitem{hudson1974wigner}
Robin~L Hudson.
\newblock When is the wigner quasi-probability density non-negative?
\newblock {\em Reports on Mathematical Physics}, 6(2):249--252, 1974.

\bibitem{kenfack2004negativity}
Anatole Kenfack and Karol {\.Z}yczkowski.
\newblock Negativity of the wigner function as an indicator of non-classicality.
\newblock {\em Journal of Optics B: Quantum and Semiclassical Optics}, 6(10):396, 2004.

\bibitem{takahashi1986wigner}
Kin'ya Takahashi.
\newblock Wigner and husimi functions in quantum mechanics.
\newblock {\em Journal of the Physical Society of Japan}, 55(3):762--779, 1986.

\bibitem{takahashi1985chaos}
Kin'ya Takahashi and Nobuhiko Sait{\^o}.
\newblock Chaos and husimi distribution function in quantum mechanics.
\newblock {\em Physical review letters}, 55(7):645, 1985.

\bibitem{gross1961structure}
Eugene~P Gross.
\newblock Structure of a quantized vortex in boson systems.
\newblock {\em Il Nuovo Cimento (1955-1965)}, 20(3):454--477, 1961.

\bibitem{pitaevskii1961vortex}
Lev~P Pitaevskii.
\newblock Vortex lines in an imperfect bose gas.
\newblock {\em Sov. Phys. JETP}, 13(2):451--454, 1961.

\bibitem{zhao2019optical}
Qiang Zhao and Jingxiang Zhao.
\newblock Optical lattice effects on shannon information entropy in rotating bose--einstein condensates.
\newblock {\em Journal of Low Temperature Physics}, 194(3):302--311, 2019.

\bibitem{chakrabarti2024quantum}
Barnali Chakrabarti, Arnaldo Gammal, ND~Chavda, and Mantile~Leslie Lekala.
\newblock Quantum-information-theoretical measures to distinguish fermionized bosons from noninteracting fermions.
\newblock {\em Physical Review A}, 109(6):063308, 2024.

\bibitem{case2008wigner}
William~B Case.
\newblock Wigner functions and weyl transforms for pedestrians.
\newblock {\em American Journal of Physics}, 76(10):937--946, 2008.

\bibitem{ojha2025phase}
Vikash~Kumar Ojha, Ramkumar Radhakrishnan, Siddharth~Kumar Tiwari, and Mariyah Ughradar.
\newblock Phase-space distributions in information theory.
\newblock {\em Pramana}, 99(1):29, 2025.

\bibitem{leonhardt2010essential}
Ulf Leonhardt.
\newblock {\em Essential quantum optics: from quantum measurements to black holes}.
\newblock Cambridge University Press, 2010.

\bibitem{lieb1978proof}
Elliott~H Lieb.
\newblock Proof of an entropy conjecture of wehrl.
\newblock {\em Communications in Mathematical Physics}, 62(1):35--41, 1978.

\bibitem{wehrl1979relation}
Alfred Wehrl.
\newblock On the relation between classical and quantum-mechanical entropy.
\newblock {\em Reports on Mathematical Physics}, 16(3):353--358, 1979.

\bibitem{cover1991network}
Thomas~M Cover and Joy~A Thomas.
\newblock Network information theory.
\newblock {\em Elements of information theory}, pages 374--458, 1991.

\bibitem{husimi1940some}
K{\^o}di Husimi.
\newblock Some formal properties of the density matrix.
\newblock {\em Proceedings of the Physico-Mathematical Society of Japan. 3rd Series}, 22(4):264--314, 1940.

\bibitem{harriman1988some}
John~E Harriman.
\newblock Some properties of the husimi function.
\newblock {\em The Journal of chemical physics}, 88(10):6399--6408, 1988.

\bibitem{appleby1999generalized}
David~M Appleby.
\newblock Generalized husimi functions: analyticity and information content.
\newblock {\em Journal of Modern Optics}, 46(5):825--841, 1999.

\bibitem{rao2004cumulative}
Murali Rao, Yunmei Chen, Baba~C Vemuri, and Fei Wang.
\newblock Cumulative residual entropy: a new measure of information.
\newblock {\em IEEE transactions on Information Theory}, 50(6):1220--1228, 2004.

\bibitem{kharazmi2023fisher}
FCE Lima.
\newblock Quantum information entropies for a soliton at hyperbolic well.
\newblock {\em Annals of Physics}, 442:168906, 2022.

\bibitem{van2014renyi}
Tim V~Erven and Peter Harremos.
\newblock R{\'e}nyi divergence and kullback-leibler divergence.
\newblock {\em IEEE Transactions on Information Theory}, 60(7):3797--3820, 2014.

\bibitem{cover1999elements}
Thomas~M Cover.
\newblock {\em Elements of information theory}.
\newblock John Wiley \& Sons, 1999.

\bibitem{mackay2003information}
David~JC MacKay.
\newblock {\em Information theory, inference and learning algorithms}.
\newblock Cambridge university press, 2003.

\bibitem{salazar2023phase}
Sa{\'u}l~JC Salazar, Humberto~G Laguna, and Robin~P Sagar.
\newblock Phase-space quantum distributions and information theory.
\newblock {\em Physical Review A}, 107(4):042417, 2023.

\bibitem{grabowski1984wehrl}
Marian Grabowski.
\newblock Wehrl-lieb's inequality for entropy and the uncertainty relation.
\newblock {\em Reports on mathematical physics}, 20(2):153--155, 1984.

\bibitem{floerchinger2021wehrl}
Stefan Floerchinger, Tobias Haas, and Henrik M{\"u}ller-Groeling.
\newblock Wehrl entropy, entropic uncertainty relations, and entanglement.
\newblock {\em Physical Review A}, 103(6):062222, 2021.

\bibitem{muller2013quantum}
Martin M{\"u}ller-Lennert, Fr{\'e}d{\'e}ric Dupuis, Oleg Szehr, Serge Fehr, and Marco Tomamichel.
\newblock On quantum r{\'e}nyi entropies: A new generalization and some properties.
\newblock {\em Journal of Mathematical Physics}, 54(12), 2013.

\bibitem{zozor2007classes}
Steeve Zozor and Christophe Vignat.
\newblock On classes of non-gaussian asymptotic minimizers in entropic uncertainty principles.
\newblock {\em Physica A: Statistical Mechanics and its Applications}, 375(2):499--517, 2007.

\bibitem{hoang2015cauchy}
Hung~Gia Hoang, Ba-Ngu Vo, Ba-Tuong Vo, and Ronald Mahler.
\newblock The cauchy--schwarz divergence for poisson point processes.
\newblock {\em IEEE Transactions on Information Theory}, 61(8):4475--4485, 2015.

\bibitem{williams1991organic}
Jack~M Williams, Arthur~J Schultz, URS Geiser, K~Douglas Carlson, Aravinda~M Kini, H~Gau Wang, Wai-Kwong Kwok, Myung-Hwan Whangbo, and James~E Schirber.
\newblock Organic superconductors—new benchmarks.
\newblock {\em Science}, 252(5012):1501--1508, 1991.

\bibitem{jerome1991physics}
Denis J{\'e}rome.
\newblock The physics of organic superconductors.
\newblock {\em Science}, 252(5012):1509--1514, 1991.

\bibitem{wehrl1978general}
Alfred Wehrl.
\newblock General properties of entropy.
\newblock {\em Reviews of Modern Physics}, 50(2):221, 1978.

\bibitem{bialynicki2006formulation}
Iwo Bialynicki-Birula.
\newblock Formulation of the uncertainty relations in terms of the r{\'e}nyi entropies.
\newblock {\em Physical Review A—Atomic, Molecular, and Optical Physics}, 74(5):052101, 2006.

\bibitem{sun2013quantum}
Guo-Hua Sun, M~Avila Aoki, and Shi-Hai Dong.
\newblock Quantum information entropies of the eigenstates for the p{\"o}schl—teller-like potential.
\newblock {\em Chinese Physics B}, 22(5):050302, 2013.

\bibitem{shi2018quantum}
Ye-Jiao Shi, Guo-Hua Sun, Farida Tahir, AI~Ahmadov, Bing He, and Shi-Hai Dong.
\newblock Quantum information measures of infinite spherical well.
\newblock {\em Modern Physics Letters A}, 33(16):1850088, 2018.

\bibitem{robertson1929uncertainty}
Howard~Percy Robertson.
\newblock The uncertainty principle.
\newblock {\em Physical Review}, 34(1):163, 1929.

\bibitem{kennard1927quantenmechanik}
Earle~H Kennard.
\newblock Zur quantenmechanik einfacher bewegungstypen.
\newblock {\em Zeitschrift f{\"u}r Physik}, 44(4):326--352, 1927.

\bibitem{sen2011statistical}
Kali~Das Sen.
\newblock {\em Statistical complexity: applications in electronic structure}.
\newblock Springer Science \& Business Media, 2011.

\bibitem{bialynicki1975uncertainty}
Iwo Bia{\l}ynicki-Birula and Jerzy Mycielski.
\newblock Uncertainty relations for information entropy in wave mechanics.
\newblock {\em Communications in Mathematical Physics}, 44:129--132, 1975.

\bibitem{coles2017entropic}
Patrick~J Coles, Mario Berta, Marco Tomamichel, and Stephanie Wehner.
\newblock Entropic uncertainty relations and their applications.
\newblock {\em Reviews of Modern Physics}, 89(1):015002, 2017.

\bibitem{solaimani2020quantum}
M~Solaimani and Shi-Hai Dong.
\newblock Quantum information entropies of multiple quantum well systems in fractional schr{\"o}dinger equations.
\newblock {\em International Journal of Quantum Chemistry}, 120(5):e26113, 2020.

\bibitem{carrillo2022shannon}
R~Santana Carrillo, Qian Dong, Guo-Hua Sun, Ramon Silva-Ortigoza, and Shi-Hai Dong.
\newblock Shannon entropy of asymmetric rectangular multiple well with unequal width barrier.
\newblock {\em Results in Physics}, 33:105109, 2022.

\bibitem{gil2022quantum}
Carlos~Ariel Gil-Barrera, Raymundo Santana~Carrillo, Guo-Hua Sun, and Shi-Hai Dong.
\newblock Quantum information entropies on hyperbolic single potential wells.
\newblock {\em Entropy}, 24(5):604, 2022.

\bibitem{santana2023quantum}
R~Santana-Carrillo, Roberto de~J.~Le{\'o}n-Montiel, Guo-Hua Sun, and Shi-Hai Dong.
\newblock Quantum information entropy for another class of new proposed hyperbolic potentials.
\newblock {\em Entropy}, 25(9):1296, 2023.

\bibitem{carrillo2021shannon}
R~Santana Carrillo, CA~Gil-Barrera, Guo-Hua Sun, M~Solaimani, and Shi-Hai Dong.
\newblock Shannon entropies of asymmetric multiple quantum well systems with a constant total length.
\newblock {\em The European Physical Journal Plus}, 136(10):1060, 2021.

\bibitem{torres2018radial}
Ariadna~J Torres-Arenas, Qian Dong, Guo-Hua Sun, and Shi-Hai Dong.
\newblock Radial position-momentum uncertainties for the infinite circular well and fisher entropy.
\newblock {\em Physics Letters A}, 382(26):1752--1759, 2018.

\bibitem{valencia2015quantum}
R~Valencia-Torres, Guo-Hua Sun, and Shi-Hai Dong.
\newblock Quantum information entropy for a hyperbolical potential function.
\newblock {\em Physica Scripta}, 90(3):035205, 2015.

\bibitem{sun2013quantum1}
Guo-Hua Sun and Shi-Hai Dong.
\newblock Quantum information entropies of the eigenstates for a symmetrically trigonometric rosen--morse potential.
\newblock {\em Physica Scripta}, 87(4):045003, 2013.

\bibitem{serrano2016information}
FA~Serrano, BJ~Falaye, and Shi-Hai Dong.
\newblock Information-theoretic measures for a solitonic profile mass schr{\"o}dinger equation with a squared hyperbolic cosecant potential.
\newblock {\em Physica A: Statistical Mechanics and its Applications}, 446:152--157, 2016.

\bibitem{deffner2025quantum}
Sebastian Deffner.
\newblock Quantum thermodynamics of gross-pitaevskii qubits.
\newblock {\em arXiv preprint arXiv:2510.12599}, 2025.

\bibitem{MariEisert2012}
A.~Mari and J.~Eisert.
\newblock Positive {W}igner functions render classical simulation of quantum computation efficient.
\newblock {\em Physical Review Letters}, 109(23):230503, 2012.

\bibitem{VeitchFerrieEmerson2012}
V.~Veitch, C.~Ferrie, D.~Gross, and J.~Emerson.
\newblock Negative quasi-probability as a resource for quantum computation.
\newblock {\em New Journal of Physics}, 14:113011, 2012.

\bibitem{BoothEmeriauKashefiMansfield2022}
R.~I. Booth, U.~Chabaud, and P.-E. Emeriau.
\newblock Contextuality and {W}igner negativity are equivalent for continuous-variable quantum measurements.
\newblock {\em Physical Review Letters}, 129(23):230401, 2022.

\bibitem{AlbarelliGenoniParisFerraro2018}
F.~Albarelli, M.~G. Genoni, M.~G.~A. Paris, and A.~Ferraro.
\newblock Resource theory of quantum non-{G}aussianity and {W}igner negativity.
\newblock {\em Physical Review A}, 98(5):052350, 2018.

\bibitem{HowardWallmanVeitchEmerson2014}
M.~Howard, J.~Wallman, V.~Veitch, and J.~Emerson.
\newblock Contextuality supplies the `magic' for quantum computation.
\newblock {\em Nature}, 510:351--355, 2014.

\bibitem{GarciaAlvarez2020}
L.~Garc{\'i}a-{\'A}lvarez, C.~Calcluth, A.~Ferraro, and G.~Ferrini.
\newblock Efficient simulability of continuous-variable circuits with large {W}igner negativity.
\newblock {\em Physical Review Research}, 2:043322, 2020.

\bibitem{ChabaudEmeriauGrosshans2021}
U.~Chabaud, P.-E. Emeriau, and F.~Grosshans.
\newblock Witnessing {W}igner negativity.
\newblock {\em Quantum}, 5:471, 2021.

\end{thebibliography}
\end{document}